\documentclass[a4paper,11pt]{article}
\newcounter{Nstates}
\usepackage[fleqn]{amsmath}
\usepackage{amssymb,amsthm}
\usepackage{forloop}

\newcounter{nn}
\newcounter{mm}
\newcounter{kk}

\def\WriteWaveFunction#1{%            Takes #1 and increments out that often  
   \setcounter{mm}{2}
   \forloop{nn}{1}{\value{nn} < \value{#1}}{
     \varphi_{\arabic{nn}}=&c_{\arabic{nn}}\ l_{\arabic{nn}}+s_{\arabic{nn}}\ \varphi_{\arabic{mm}}\\
     \stepcounter{mm}
   }
  \varphi_{\arabic{#1}}=&l_{\arabic{#1}}
}

\def\WriteDensityMatrix#1{% Takes #1 and increments out that often  
   \setcounter{mm}{2}
   \forloop{nn}{1}{\value{nn} < \value{#1}}{
     \varphi_{\arabic{nn}}\varphi_{\arabic{nn}}^\dagger=&\frac{1}{2}+\frac{1}{2}(l_{\arabic{nn}}l_{\arabic{nn}}^\dagger-\varphi_{\arabic{mm}}\varphi_{\arabic{mm}}^\dagger) (\cos(2\theta_{\arabic{nn}})+J_{\arabic{nn}}\sin(2\theta_{\arabic{nn}}))\\
     \stepcounter{mm}
   }
  \varphi_{\arabic{#1}}\varphi_{\arabic{#1}}^\dagger=&l_{\arabic{#1}}l_{\arabic{#1}}^\dagger
}

\def\WriteBasis#1{%            Takes #1 and increments out that often  
   \{
   \forloop{nn}{1}{\value{nn} < \value{#1}}{
     l_{\arabic{nn}},
   }
  l_{\arabic{#1}}\}
}

\def\WriteBorn#1{%            Takes #1 and increments out that often  
  \forloop{nn}{1}{\value{nn} < \value{#1}}{
    P(\arabic{nn})=&(
    \forloop{mm}{1}{\value{mm} < \value{nn}}{
       s_{\arabic{mm}}\ 
     }
     c_{\arabic{nn}})^2\\
   }
   P(\arabic{#1})=&(
   \forloop{mm}{1}{\value{mm} < \value{#1}}{
       s_{\arabic{mm}}\ 
     }
    )^2
  }

\def\WriteConditional#1{%            Takes #1 and increments out that often  
  \setcounter{mm}{2}
  \forloop{nn}{1}{\value{nn} < \value{#1}}{
    P( \arabic{nn} | (\arabic{nn} \mathrm{\ or\ above}))&=\frac{1}{2}+\frac{1}{2}\cos(2\theta_{\arabic{nn}})\ \ \ \ \ 
    P( (\arabic{mm} \mathrm{\ or\ above}) | (\arabic{nn} \mathrm{\ or\ above}))= \frac{1}{2}-\frac{1}{2}\cos(2\theta_{\arabic{nn}})\\
    \stepcounter{mm}
  }
  P( \arabic{#1} | (\arabic{#1} \mathrm{\ or\ above}))&=1
}

\def\WriteBornConditional#1{%            Takes #1 and increments out that often  
  \forloop{nn}{1}{\value{nn} < \value{#1}}{
    P(\arabic{nn})=
    \setcounter{kk}{2}
    \forloop{mm}{1}{\value{mm} < \value{nn}}{
      & P( (\arabic{kk} \mathrm{\ or\ above}) | (\arabic{mm} \mathrm{\ or\ above}))*\\ 
      \stepcounter{kk}
    }
     & P( \arabic{nn} | (\arabic{nn} \mathrm{\ or\ above}))\\
   }
    P(\arabic{nn})=
    \setcounter{kk}{2}
    \forloop{mm}{1}{\value{mm} < \value{nn}}{
      & P( (\arabic{kk} \mathrm{\ or\ above}) | (\arabic{mm} \mathrm{\ or\ above}))*\\ 
      \stepcounter{kk}
    }
     & P( \arabic{nn} | (\arabic{nn} \mathrm{\ or\ above}))
  }

\usepackage{longtable}
\usepackage{graphicx}
\usepackage{bm}% bold math
\usepackage{footmisc}
\usepackage{afterpage}
\usepackage{float}
\usepackage{lscape}

\usepackage[square,comma,sort&compress,numbers]{natbib}

\usepackage[dvipsnames]{xcolor}
\definecolor{gray75}{gray}{0.4}
\usepackage{fullpage}
\usepackage{etex}
\usepackage[utf8]{inputenc}
\usepackage[T1]{fontenc}
\usepackage{lmodern}
\usepackage[nolist,nohyperlinks]{acronym}
\usepackage{booktabs}
\usepackage{hepunits}
\usepackage{units}

\usepackage{multirow}

\usepackage{epigraph}
\epigraphsize{\small\itshape}
\usepackage{csquotes}

\usepackage{slashed}
\usepackage[section]{easy-todo}
\usepackage{caption}
\usepackage{subcaption}
\usepackage{setspace}
\usepackage{notoccite}
\usepackage{listings}
\usepackage{enumerate}
\usepackage[hyperfootnotes=false,pdfpagelabels]{hyperref}  % backref linktocpage pagebackref

\hypersetup{%
    colorlinks=true, linktocpage=true, pdfstartpage=3, pdfstartview=FitV,%
    breaklinks=true, pdfpagemode=UseNone, pageanchor=true, pdfpagemode=UseOutlines,%
    plainpages=false, bookmarksnumbered, bookmarksopen=true, bookmarksopenlevel=1,%
    hypertexnames=true, pdfhighlight=/O,%nesting=true,%frenchlinks,%
   urlcolor=gray75, linkcolor=gray75, citecolor=gray75,
     pdftitle={Collapse and Euler},%
     pdfauthor={Leonardo Antunes Pedro},%
}

\usepackage{mciteplus}

\newcommand*{\bea}{\begin{eqnarray}}
\newcommand*{\eea}{\end{eqnarray}}
\newcommand*{\be}{\begin{equation}}
\newcommand*{\ee}{\end{equation}}

\newcommand*{\tr}{\mathrm{tr}}
\newcommand*{\diag}{\mathrm{diag}}

\newcommand{\bma}{\begin{pmatrix}}
\newcommand{\ema}{\end{pmatrix}}

\graphicspath{{./Images/}}

\renewcommand{\textflush}{flushepinormal}

\theoremstyle{plain}% default

\newtheorem*{rmk*}{Note}

\theoremstyle{definition}

\newtheorem*{defn*}{Definition}

\theoremstyle{remark}

\makeatletter
\renewcommand{\@epitext}[1]{
\itshape \begin{minipage}{\epigraphwidth}\begin{\textflush} #1
\end{\textflush}\end{minipage}\vspace{1ex}}

\makeatother
\title{Relating the wave-function collapse with Euler's formula, with applications to Classical Statistical Field Theory}
\author{Leonardo Pedro\\
  leonardo@cftp.tecnico.ulisboa.pt}
\date{\today}
\begin{document}
\maketitle

\begin{abstract}    
    
One attractive interpretation of quantum mechanics is the ensemble interpretation, where Quantum Mechanics merely describes a statistical ensemble of objects and not individual objects. But this interpretation does not address why the wave-function plays a central role in the calculations of probabilities, unlike most other interpretations of quantum mechanics.
On the other hand, Classical Statistical Field Theory suffers from severe mathematical inconsistencies (specially for Hamiltonians which are non-polynomial in the fields, e.g. General relativistic statistical field theory). We claim that both problems are related to each other and we propose a solution to both.

We prove: 1) the wave-function is a parametrization of any probability distribution of a statistical ensemble: there is a surjective map from an hypersphere to the set of all probability distributions;\\
2) for a quantum system defined in a 2-dimensional real Hilbert
space, the role of the (2-dimensional real) wave-function is identical
to the role of the Euler's formula in engineering, while the collapse of the  wave-function is identical to selecting the real part of a complex number;\\
3) the collapse of the wave-function of any quantum system is a recursion of collapses of 2-dimensional real wave-functions;\\
4) the wave-function parametrization is key in the mathematical definition of Classical Statistical Field Theory we propose here. The same formalism is applied to Quantum Yang-Mills theory and Quantum Gravity in another article. 

%The wave-function plays a central role because it is a good parametrization that allows us to represent a  group of transformations using linear transformations of the hypersphere. It is precisely the fact that the hypersphere is not the phase-space of the theory that implies the collapse of the wave-function. Without collapse, the wave-function parametrization would be inconsistent.
\end{abstract}

\section{Introduction}

The mainstream literature on quantum mechanics claims that 1) the mathematical formalism of quantum mechanics is clear, however 2) the interpretation of the mathematical formalism is far from clear because the quantum phenomena defies our everyday view of the world. In summary, that there is no technical problem with quantum mechanics, only an interpretation problem. This is illustrated by Richard Feynman saying that: ``I think I can safely say that nobody understands Quantum Mechanics.''

In this article we will show that the interpretation of quantum mechanics is clear and the quantum phenomena does not defy our everyday view of the world, so 2) is false. We also propose a solution to the severe mathematical inconsistencies of Classical Statistical Field Theory here (specially for Hamiltonians which are non-polynomial in the fields, e.g. General relativistic statistical field theory) and of Field Theories in general elsewhere~\cite{pedro_1442442}, so 1) is also false.
%In another article, we will show that the definition of gauge theories within the formalism of quantum mechanics can be improved to be compatible with a non-deterministic time-evolution~\cite{pedro_1442442}, so 1) is also false Note that a mathematically consistent definition of a Quantum Yang-Mills theory in a 4-dimensional space-time (such as the Standard Model of Particle Physics) was not achieved so far~\cite{prize}, and as a consequence the mainstream definitions of a Quantum Yang-Mills theory are based on a classical Lagrangian where the time-evolution is deterministic.
It is usually argued in the literature that the mathematical inconsistencies only affect Field Theory and not Quantum Mechanics. But this argument is misleading because if every quantum system is made of fundamental Particles, then Quantum Mechanics must at least be consistent with Quantum Field Theory. Moreover, the Hamiltonian formulation of conservative classical mechanics cannot be extended in a
straightforward manner to time-dependent classical mechanics because the symplectic form is not invariant under time-dependent transformations; the most obvious way out is to formulate non-relativistic time-dependent mechanics as a particular field
theory whose configuration space is a fibred manifold over a time axis~\cite{gaugemechanics2,gaugemechanics}.

We will show in this article that the mathematical inconsistencies of Field Theory do have consequences for the interpretation of the quantum mechanics of electrons and atoms and the interpretation of quantum mechanics in general. 
In summary, we argue that in fact there is a technical problem with the definition of Field Theories (with a solution), and there is no interpretation problem with Quantum Mechanics (beyond what is reasonable to expect for any theory of Physics).

Often quantum mechanics is interpreted as providing the probabilities of transition  between different states of an individual system. This transition happens upon measurement, any measurement. The state of the system is defined by the wave-function which collapses to a different wave-function upon measurement.

This raises a number of interpretation problems as to what do we mean by state of a system. If the state before measurement is A, but after measurement the state is B, what is then the state during the measurement: A and/or B or something else? Strangely, despite that we don't know what happens during the measurement, we know very well the transition probabilities---because once we assume that the state of the system is defined by the wave-function, there are not (many) alternatives to the Born's rule defining the probabilities of transition as a function of the wave-function~\cite{gleason1957,*2013gleason}.

The ensemble interpretation where Quantum Mechanics is merely a formalism of statistical physics, describing a statistical ensemble of systems~\cite{1992PhR...210..223H,ballentine} avoids these interpretation problems: the state of the ensemble is unambiguously the probability distribution for the states of an individual system---as in classical statistical mechanics\footnote{Note that the notion of probability distribution is not free from interpretation problems~\cite{gillies2000philosophical}, but we believe these are intrinsic to all applications of probabilities.}. Then, there are several possible states of the ensemble, these different states are related by symmetry transformations---such as a translation in space-time or a rotation in space. Note that in classical statistical mechanics, the symmetry transformations are deterministic (unlike in Quantum Mechanics), i.e. they transform deterministic ensembles into deterministic ensembles (by deterministic ensemble we mean that all systems in the statistical ensemble are in the same state of the classical phase space). 
    
Since we can always define a wave-function by taking the square-root of the probabilities (see Section~\ref{sec:prob}), the Koopman-von Neumann version of classical statistical mechanics~\cite{Sudarshan1976} defines classical statistical mechanics as a particular case of quantum mechanics where the algebra of operators is necessarily commutative (because the symmetry transformations are deterministic).

Quantum mechanics in the ensemble interpretation, generalizes classical statistical mechanics by allowing symmetry transformations of the statistical ensemble of systems to be non-deterministic. For instance, the probability clock~\cite{NOOM}[see also our Section~\ref{sec:euler}] involves a non-deterministic symmetry transformation. In classical statistical mechanics any non-deterministic transformation is an external foreign element to the theory, this is unnatural for a statistical theory. Thus Quantum Mechanics in the ensemble interpretation is a natural and unavoidable generalization of classical statistical mechanics.

%The set of possible states of the ensemble can be parametrized in a meaningful way with respect to the physical transformations.

However, the ensemble interpretation does not address the question why the wave-function plays a central role in the calculation of the probability distribution, unlike most other interpretations of quantum mechanics. By being compatible with most (if not all) interpretations of Quantum Mechanics, the ensemble interpretation is in practice a common denominator of most interpretations of Quantum Mechanics. It is useful, but it is not enough.
For instance, the ensemble interpretation does not give any explanation as to why  it looks like the electron's wave-function interferes with itself in the double-slit experiment~\cite{feynmandoubleslit,bookyoung}[see also Section~\ref{sec:doubleslit}]---that would imply that the wave-function describes (in some sense) an individual system.  Also, the ensemble interpretation does not explain the role of Quantum Statistical Mechanics and the associated density matrix in the measurement process: due to the wave-function's collapse, the off-diagonal part of the density matrix is always set to zero, while the diagonal part of the density matrix containing the probabilities of transition is preserved---in a basis where the operators (corresponding to the measurable properties of the system) are diagonal (see Section~\ref{sec:density}). Also, if a black-hole erases most information about an object that comes inside of it by turning this information to random, then it is not obvious how the symmetry of translation in time is conserved in the ensemble interpretation (see Section~\ref{sec:informationparadox}).
Moreover, the most prominent advocates of the ensemble interpretation were dissatisfied with the complementarity of position and momentum~\cite{epr,revballentine}, convincing themselves and others that the complementarity of position and momentum could not be satisfactorily explained by the ensemble interpretation alone.

In this article we show that the wave-function can be described as a multi-dimensional generalization of Euler's formula, and its collapse as a generalization of taking the real part of Euler's formula. The wave-function is nothing else than one possible parametrization of any probability distribution; the parametrization is a surjective map from  an hypersphere to the set of all possible probability distributions. The fact that the hypersphere is a surface of constant radius reflects the fact that the integral of the probability distribution is always $1$. Two wave-functions are always related by a rotation of the hypersphere, which is a linear transformation and it preserves the hypersphere. 
It is thus a good parametrization which allows us to represent a group of symmetry transformations using linear transformations of the hypersphere. These symmetry transformations may be deterministic, thus quantum mechanics is a generalization of classical statistical mechanics (but not of probability theory).

This is ironic, since Feynman described Euler's formula as ``our jewel'' while the wave-function collapse certainly contributed for him to say ``I think I can safely say that nobody understands Quantum Mechanics.'' Besides the irony, the fact that the wave-function is nothing else than one possible parametrization of any probability distribution means that the  wave-function collapse is a feature of all random phenomena. 

The above fact implies that alternatives to Quantum Mechanics motivated by a dissatisfaction with either the complementarity of position and momentum~\cite{revballentine},
or the wave-function collapse~\cite{Weinberg:2016uml}, may also feature complementarity of observables (possibly other than position and momentum) and wave-function collapse,
once a parametrization with a wave-function is applied. The physical question is how the physical transformations affect the ensemble (and thus the wave-function), in particular whether there are viable alternative theories to Quantum Mechanics where the time-evolution of the statistical ensemble is deterministic~\cite{automaton} or at least it is a stochastic process~\cite{nelsonreview,WeinbergLindblad,stochasticliouville}; in such a case  there would be other alternative parametrizations which also allows us to use methods of group theory and do not involve a wave-function and its collapse.

But the above fact also implies that the wave-function collapse is not provoked by the interaction with the environment\footnote{In a measurement there is always an interaction with the environment, therefore the environment necessarily affects the ensemble and it is possible that decoherence occurs. But such phenomena will be accounted for by the time-evolution of the ensemble. Note that a continuous (repeated) quantum measurement is a model of decoherence and thus decoherence does not avoid by itself the wave-function collapse~\cite{mensky2000quantum,haroche2006exploring}. In case we opt for a model of decoherence which avoids collapse, then we are necessarily dealing with an alternative to Quantum Mechanics, such case was discussed above.}; the wave-function collapse does not emerge from some particular cases of classical statistics~\cite{fermionpath3}; Quantum Mechanics is \emph{not} a generalization of the concept of probability algebra from commutative to non-commutative algebra~\cite{2000JMP....41.3556S}; and thus quantum computation/information is not fundamentally different from classical computation/information\footnote{Different computers always have different properties, for instance different logic gates may enhance the performance of different algorithms~\cite{information}. But neglecting performance, the quantum bits can be constructed using classical bits and quantum logic gates can also be constructed using classical logic gates, with the Hadamard transform as an example.}. The wave-function is a possible parametrization for any theory of Statistics, including Statistical Physics.

This is comforting, since it is consistent with the empirical facts that Quantum Mechanics applies to a very wide range of physical systems, from the Hydrogen atom, a neutron star or the Universe; and that the collapse occurs upon measurement, any measurement. This also opens the door into applying the wave-function parametrization not just to quantum mechanics and quantum statistical mechanics, but also to quantum field theory or even to other problems involving statistics other than quantum physics~\cite{bertinstatistical}. For instance, sequential systems are a framework for machine learning that shares several features with Quantum Mechanics~\cite{qmm, jaeger, NOOM}, (more) application of quantum methods to sequential systems thus seems straightforward. Quantum Reinforcement Learning shows superior performance in computer simulations and recently, it was empirically tested on human decision-making~\cite{qrl}. Other applications of quantum methods in statistics, either did not use the wave-function parametrization\footnote{Analogies with the wave-function were made but unitarity was not preserved~\cite{baezreview,quantumforstat,small2011hilbert,hagen2009path}}; or they considered only deterministic dynamics~\cite{rkhs,marinho,quantumsimulation}.

In Section~\ref{sec:prob}, we show that the wave-function is one possible parametrization of any probability distribution\footnote{Such parametrization is also implicitly used in the literature based on the Koopman-von Neumann version of classical statistical mechanics~\cite{Sudarshan1976}.}; in Section~\ref{sec:density} we discuss the difference between the wave-function parametrization and the density matrix parametrization of Gleason's theorem; in Sections~\ref{sec:symmetries} and~\ref{sec:deterministic} we discuss symmetry transformations and deterministic transformations; 
%we  review the representation of an algebra of events in a real Hilbert space (which will be useful for the next sections), and we argue that quantum mechanics is \emph{not} a non-commutative version of probability theory; 
in Section~\ref{sec:euler} we describe the relation between Euler's formula and the parametrization of a probability distribution by a real wave-function; in Sections~\ref{sec:sg} and ~\ref{sec:informationparadox} we describe the Stern-Gerlach experiment and Black Hole Information paradox using the Euler's formula and the ensemble interpretation; in Sections~\ref{sec:mapN} and~\ref{sec:map} we describe the parametrization of a probability distribution by a real wave-function (i.e. for a finite and generic number of states, respectively); in Section~\ref{sec:complex} we address complex and quaternionic wave-functions; in Section~\ref{sec:stochastic} we distinguish the time-evolution in quantum mechanics from a stochastic process; in Section~\ref{sec:timetranslation} we show that the the time-evolution in quantum mechanics is a stochastic process if and only if it is deterministic; in Section~\ref{sec:irreversible} we discuss how the concept of (ir)reversible processes from thermodynamics applies to non-deterministic symmetry transformations; in Section~\ref{sec:complete} we show that quantum mechanics is a complete description of physical reality in the sense of Einstein-Podolsky-Rosen; in Section~\ref{sec:relativistictheory} we show that any deterministic theory compatible with
relativistic Quantum Mechanics necessarily respects relativistic causality; in Section~\ref{sec:deterministictheory} we build an explicit example of a deterministic theory compatible with relativistic Quantum Mechanics; in Section~\ref{sec:doubleslit} we describe the double-slit experiment using the ensemble interpretation; in Section~\ref{sec:bell} we argue that our results imply that the Bell inequalities merely establish a difference between Quantum Mechanis and a theory where time-evolution is a stochastic process, unlike what is often claimed in the literature; in Section~\ref{sec:constraints} we will define constrained systems and the conditioned probability in particular; in Section~\ref{sec:sft} we will define an appropriate time-evolution to statistical field theory using constraints;
%in Section~\ref{sec:gns} we provide another proof (shorter but less intuitive) of the fact that  the wave-function is a parametrization of any probability distribution, and we also discuss why this fact was missed by many people; 
we conclude in Section~\ref{sec:conclusion}.

Note that in this paper, the Hilbert space is always considered to be
a separable Hilbert space~\cite{pct} and unless otherwise stated it is
a real Hilbert space. 
%We strongly oppose this view\footnote{Note that unexpected phenomena can always be considered mystical in the short term but not many decades later.}, there is no way in which a non-deterministic time-evolution can defy our everyday view of the world\footnote{Certainly, a non-deterministic time-evolution is compatible with the theory of classical information. The concept ``quantum information'' has nothing to do with ``science''.}.  In fact, it is the other way around: our everyday experience of the world involves a non-deterministic time-evolution; also from the point of view of mathematical complexity, many deterministic equations of motion may be extremely complex to solve while the time-evolution in quantum mechanics is a linear transformation~\cite{pedro_epr}.

%; in Section~\ref{sec:issues} we will discuss the complementarity of position and momentum and the case of a non-instantaneous collapse of the wave-function.

\section{The wave-function is a parametrization of any probability distribution}
%\section{Quantum Mechanics is \emph{not} a non-commutative version of probability theory}
\label{sec:prob}

The representation of an algebra of events in a real Hilbert space uses projection-valued
measures~\cite{realpoincare,Oppio:2016pbf,realoperatoralgebras,moretti2013}.
A probability space consists of three parts: the phase space (which is the set of possible states of a system); the set of events  where each event is a subset of the set of possible states;
and a probability distribution (also named a probability measure) which assigns a probability to each event.

The notion of probability is somewhat ambiguous~\cite{gillies2000philosophical}, but it is useful to relate complex random phenomena with a simple standard random process.
That the probability of an event is $5/567$ means that the likelihood of our event is the same as the likelihood of picking one red ball out of a bag with 567 balls where 5 balls are red (standard random process).
If the probability is a real number not rational, we can approximate any real probability by a rational number with infinitesimal error
because the rational numbers are dense in the reals, therefore the relation to a simple standard random process is still possible.

A projection-valued measure assigns a self-adjoint projection operator of a real Hilbert space to each event, in such a way that the boolean algebra of events is represented by the commutative algebra of projection operators. Thus, intersection/union of events is represented by products/sums of projections, respectively.

The state of the ensemble is a linear functional which assigns a probability to each projection. We now show that the wave-function is one possible parametrization of any probability distribution. That is, for any state of the ensemble, there is a wave-function such that the probability of any event is given by the Born rule. This result is not surprising since in principle, we should always be able to define a wave-function by taking the square-root of the probabilities. This parametrization is also implicitly used in the literature based on the Koopman-von Neumann version of classical statistical mechanics~\cite{Sudarshan1976}. However we want to apply this parametrization beyond classical statistical mechanics to general Quantum Mechanics including non-deterministic transformations and the wave-function collapse, so we need a solid explicit proof of this result to clarify the limits of applicability of the parametrization~\footnote{We couldn't found a solid explicit proof of the wave-function parametrization.}. Our proof is robust because it is based on the GNS construction\footnote{For more information on the Gelfand-Naimark-Segal (GNS) construction see Ref.~\cite{formalisms} for the case of a complex algebra and Ref.~\cite{realoperatoralgebras} for the case of a real algebra}. The proof follows.

The algebra of projection-valued measures $A$ associated to a measurable space $X$ is a commutative real C* algebra.
The expectation value $E$ is a positive linear functional. The expectation value allows us to define the bilinear form:
\begin{align}<a,b>=E(a b)\end{align}

where $a,b\in A$. This bilinear form is not yet an inner-product, since it is only positive semi-definite. However, the set of projections with null expectation value $I_E$ is a linear subspace of $A$.
Thus, the completion of the quotient $A/I_E$  is an Hilbert space (with inner product given by the above bilinear form, this is the GNS construction). The vector $v_0$ corresponding to the identity $1$ of the algebra is a cyclic vector. The projection-valued measures correspond to the projection of $v_0$ in a corresponding region of the space $X$.

Thus the Hilbert space corresponding to $A/I_E$ is the space of square-integrable functions in the region of $X$ where the expectation value is not null.
We need to go now beyond the GNS construction and consider instead the Hilbert space of square-integrable functions in all $X$, since we still have that 
\begin{align}<v_0,P_Y v_0>=E(P_Y)\end{align} where $P_Y$ is a projection in a subset $Y\subset X$; but in this case $v_0$ is not necessarily a cyclic vector, since the projections of $v_0$ in subspaces of $X$ do not necessarily constitute a basis of the Hilbert space.

\section{Gleason's theorem and a non-commutative generalization of probability theory}
\label{sec:density}

Since the boolean algebra of events is commutative, there is a basis where all the corresponding projections are diagonal. This leaves room for a non-commutative generalization of probability theory, since the state of the ensemble could also assign a probability to non-diagonal projections, these non-diagonal projections would generate a non-commutative algebra~\cite{2000JMP....41.3556S}. 

Consider for instance the projection $P_X$ to a region of space $X$ and a  projection $U P_P U^\dagger$ to a region of momentum $P$, where $P_X$ and $P_P$ are diagonal in the same basis.
The projections $P_X$ and $U P_P U^\dagger$ are related by a Fourier transform $U$ and thus are diagonal in different basis and do not commute (they are complementary observables).
Since we can choose to measure position or momentum, it seems that Quantum Mechanics is a  non-commutative generalization of probability theory~\cite{2000JMP....41.3556S}.

But due to the wave-function collapse, Quantum Mechanics is not a non-commutative generalization of probability theory despite the appearances:
the measurement of the momentum is only possible if a physical transformation of the statistical ensemble also occurs.
Suppose that $E(P_X)$ is the probability that the system is in the region of space $X$, for the state of the ensemble $E$ diagonal (i.e. verifying $E(O)=0$ for operators $O$ with null diagonal).
Then we define:
\begin{align}
  E_U(D)= E(U D U^\dagger)\\
  E_U(O)=0\label{eq:collapse}
\end{align}

Where $D$ is a diagonal operator and $O$ is an operator with null diagonal.
The equation~(\ref{eq:collapse}) is due to the wave-function collapse.
Thus $E_U(P_P)=E(U P_P U^\dagger)$ is the probability that the system is in the region of momentum $P$, for the state of the ensemble $E_U$.
But the ensembles $E$ and $E_U$ are different, there is a physical transformation relating them.

Without collapse, we would have $E_U(O)=E(U O U^\dagger)\neq 0$ for operators $O$ with null-diagonal and we could talk about a common state of the ensemble $E$ assigning probabilities to a non-commutative algebra. But the collapse keeps Quantum Mechanics as a standard probability theory, even when complementary observables are considered. We could argue that the collapse plays a key role in the consistency of the theory, as we will see below.

At first sight, our result that the wave-function is merely a parametrization of that any probability distribution, resembles Gleason's theorem~\cite{gleason1957,*2013gleason}.
However, there is a key difference: we are dealing with commuting projections and consequently with the wave-function, while
Gleason's theorem says that any probability measure for all \emph{non-commuting} projections defined in a Hilbert space (with dimension $\geq 3$) can be parametrized by a density matrix.
Note that a density matrix includes mixed states, and thus it is more general than a pure state which is represented by a wave-function.

We can check the difference in the 2-dimensional real case. Our result is that there is always a wave-function $\Psi$ such that $\Psi^2(1)=\cos^2(\theta)$ and $\Psi^2(2)=\sin^2(\theta)$ for any $\theta$.

However, if we consider non-commuting projections and a diagonal constant density matrix $\rho=\frac{1}{2}$, then we have: 
\begin{align}
\begin{cases}
Tr(\rho \left[\begin{smallmatrix} 1 & 0\\ 0 & 0 \end{smallmatrix}\right])=\frac{1}{2}\\
Tr(\rho \frac{1}{2}\left[\begin{smallmatrix} 1 & 1\\ 1 & 1 \end{smallmatrix}\right])=\frac{1}{2}
\end{cases}
\end{align}
Our result implies that there is a pure state, such that:
\begin{align}
Tr(\rho \left[\begin{smallmatrix} 1 & 0\\ 0 & 0 \end{smallmatrix}\right])=\frac{1}{2}
\end{align}

(e.g. $\rho=\frac{1}{2}\left[\begin{smallmatrix} 1 & 1\\ 1 & 1 \end{smallmatrix}\right]$)

And there is another possibly different pure state, such that:
\begin{align}
Tr(\rho \frac{1}{2}\left[\begin{smallmatrix} 1 & 1\\ 1 & 1 \end{smallmatrix}\right])=\frac{1}{2}
\end{align}
(e.g. $\rho=\left[\begin{smallmatrix} 1 & 0\\ 0 & 0 \end{smallmatrix}\right]$)

But there is no $\rho$ which is a pure state, such that:
\begin{align}
\begin{cases}
Tr(\rho \left[\begin{smallmatrix} 1 & 0\\ 0 & 0 \end{smallmatrix}\right])=\frac{1}{2}\\
Tr(\rho \frac{1}{2}\left[\begin{smallmatrix} 1 & 1\\ 1 & 1 \end{smallmatrix}\right])=\frac{1}{2}
\end{cases}
\end{align}

On the other hand, Gleason's theorem implies that there is a $\rho$ which is a mixed state, such that :
\begin{align}
\begin{cases}
Tr(\rho \left[\begin{smallmatrix} 1 & 0\\ 0 & 0 \end{smallmatrix}\right])=\frac{1}{2}\\
Tr(\rho \frac{1}{2}\left[\begin{smallmatrix} 1 & 1\\ 1 & 1 \end{smallmatrix}\right])=\frac{1}{2}
\end{cases}
\end{align}

Gleason's theorem is relevant if we neglect the wave-function collapse, since it attaches a unique density matrix to non-commuting operators. However, the wave-function collapse affects differently the density matrix when different non-commuting operators are considered, so that after measurement the density matrix is no longer unique. In contrast, the wave-function collapse plays a key role in the wave-function parametrization of a probability distribution.

Another difference is that our result applies to standard probability theory, while Gleason's theorem applies to a non-commutative generalization of probability theory.

\section{Symmetries and unitary representations }
\label{sec:symmetries}

In general, the probability distribution for the state of a system is a function of an homogeneous space for a group. The homogeneous space is defined as a topological space where the group acts transitively. For instance, in the case of time-evolution the probability distribution is a function of a point in a real line which in turn is an homogeneous space for the group of translations in time.

In the case of the wave-function parametrization, the probability distribution for the state of a system is a function of the space of wave-functions,  which is a multi-dimensional sphere and thus it is an homogeneous space for the group of rotations. This means that for any (normalized) wave-functions $\psi, \phi$, there is a unitary operator $U$ such that $\psi=U\phi$. There is a probability distribution associated to each wave-function and also an elementary event associated to each wave-function, as we will see in the following.

%This leaves us in a good position to define an action of any symmetry group on the wave-function. We show in the following that the space of wave-functions is the homogeneous space for the biggest symmetry group possible.

We can choose a reference wave-function $\phi$ and move the unitary matrix $U$ from $\psi=U\phi$ to the projection operators $P_A$ corresponding to the event $A$, i.e. $U\phi\to \phi$ and 
$P_A\to U^{\dagger} P_A U$.
The choice of the reference wave-function is arbitrary, since a unitary transformation $V$ acting as $\phi\to V \phi$ and $U^{\dagger} P_A U\to V U^{\dagger} P_A U V^\dagger$ conserves the probability distribution. 

For discrete probability distributions, the projection corresponding to the elementary event $n$ can be written as $P_n=\psi_n \psi_n^\dagger$ and the set of wave-functions $\{\psi_n\}$ is an orthonormal basis of the Hilbert space. 

Consider now the square of the inner product of 2 wave-functions: $\phi^\dagger\psi\psi^\dagger\phi$. 
We can write  $\psi\psi^\dagger=U P_1 U^\dagger$ and $\phi\phi^\dagger=V P_1 V^\dagger$ and then $\phi^\dagger\psi\psi^\dagger\phi=((U^\dagger V)_{11})^2$. Thus, the square of the inner product of 2 wave-functions equals the probability of the event 1 given by the wave-function obtained by applying $U^\dagger V$ to the reference wave-function $\psi(n)=\delta_{n1}$.

By definition, the inner product is invariant under the transformation $\psi\to T\psi$ and
$\phi\to T\phi$, where $T$ is  a linear isometry. The question we make now is: under which transformations are left invariant all the squares of inner products of 2 wave-functions?

If all squares of the inner product of 2 wave-functions are left invariant under a transformation $(T)$, then (at least for discrete probability distributions) both wave-functions $\psi$ and $T(\psi)$ can be associated to the same probability distribution and to the same elementary event. That is, the wave-function parametrization of a probability distribution is not necessarily unique and it is related to another parametrization by the transformation $T$. The transformation $T$ is called a symmetry (in the context of Wigner's theorem).

Wigner's theorem~\cite{2014PhLA..378.2054G,Ratz1996,*wignertheorem} implies that a symmetry is necessarily a linear isometry. Thus a symmetry also conserves the wave-function parametrization for continuous probability distributions, because it is a linear isometry. In the case of a group of symmetries, the transformations must be invertible. Since an invertible isometry is a unitary transformation, the action of a group of symmetries is necessarily linear and unitary.

In conclusion, the reference wave-function is determined up to a symmetry transformation, which is a linear isometry. This implies that the action of a group of symmetries on the reference wave-function is linear  and unitary. Moreover, it implies that the reference wave-function can be chosen to be concentrated (around) a single point of the phase-space (see also Section~\ref{sec:constraints}).
%and any probability can be described as a probability of transition between (neighborhoods of) points of the phase-space.

Note that there is not necessarily a group action of a symmetry group on the probability distribution (for the state of a system) itself. We address in Section~\ref{sec:timetranslation} when such action on the probability distribution exists and when it does not exist.

\section{Deterministic transformations}
\label{sec:deterministic}

Crucially, the symmetry transformations include all the deterministic transformations, which will be defined in the following. Thus the symmetry transformations are a generalization of the deterministic transformations.

A deterministic transformation acts as $E(P_A)\to E(P_B)$ where $A,B$ are events and $P_A$ is a projection operator, for any expectation functional $E$ and event $A$. When the probability is concentrated in the neighborhood of a single outcome (say $A$), we have effectively a deterministic case and this transformation ($A\to B$) conserves the determinism, thus it is a deterministic transformation.

Note that above, $P_A$ and $P_B$ necessarily commute. On the other hand, if the transformation is such that $E(P_A)\to E(U P_A U^\dagger)$ where $U$ is a unitary operator and $P_A$ and 
$U P_A U^\dagger$ do not commute, then the transformation cannot be deterministic. Consider
the discrete case with $E(P_n)$ given by $Tr(P_m P_n)=\delta_{mn}$ up to a normalization factor, for instance. Then $Tr(P_mP_n)\to Tr(P_mU P_n U^\dagger)=U^2_{nm}$. If the transformation would be deterministic, then necessarily $U^2_{nm}=\delta_{kn}$ for some $k=f(n)$ dependent on $n$, and so $U P_n U^\dagger=P_{l}$ with $l=f^{-1}(n)$ would commute with $P_n$. 

We conclude that a transformation $U$ is deterministic if and only if $P_A$ and $U P_A U^\dagger$ commute for all events $A$. Thus, the complementarity of two observables (e.g. position and momentum) is due to the random nature of the symmetry transformation relating the two observables. This clarifies that probability theory has no trouble in dealing with non-commuting observables, as long as the collapse of the wave-function occurs. Note that Quantum Mechanics is not a generalization of probability theory, but it is definitely a generalization of classical mechanics since it involves non-deterministic symmetry transformations. For instance, the time evolution may be non-deterministic unlike in classical mechanics.

\section{Euler's formula for the probability clock}
\label{sec:euler}

The previous sections established that the ensemble interpretation is self-consistent. However, the ensemble interpretation does not address the question why the wave-function plays a central role in the calculation of the probability distribution, unlike most other interpretations of quantum mechanics. By being compatible with most (if not all) interpretations of Quantum Mechanics, the ensemble interpretation is in practice a common denominator of most interpretations of Quantum Mechanics. It is useful, but it is not enough.

In this and in the following sections we will show that the wave-function is nothing else than one possible parametrization of any probability distribution.  The wave-function can be described as a multi-dimensional generalization of Euler's formula, and its collapse as a generalization of taking the real part of Euler's formula. The wave-function plays a central role because it is a good parametrization that allows us to represent a  group of transformations using linear transformations of the hypersphere. It is precisely the fact that the hypersphere is not the phase-space of the theory that implies the collapse of the wave-function. Without collapse, the wave-function parametrization would be inconsistent.

Suppose that we have an oscillatory motion of a ball, with position $x=\cos(t)$ and we want to make a translation in time\footnote{Here the time $t$ is merely a parameter and not necessarily the physical time. We call this arbitrary parameter ``time'' $t$ for pedagogical reasons, as is common practice in the pedagogical literature about periodic functions (e.g. about Fourier analysis).}, $\cos(t)\to \cos(t+a)$.
This is a non-linear transformation. However, if we consider not only the position but also the velocity of the ball, we have the ``wave-function'' given by the Euler's formula $q(t)=e^{it}$ and $x$ is the real part of $q$. Then, a translation is represented by a rotation $q(t+a)=e^{ia} q(t)$. To know $x$ after the translation, we need to take the real part of the wave-function $e^{ia} q(t)$, \emph{after} applying the translation operator.

Of course, $\cos(t)$ is not positive and so it has nothing to do with probabilities.
However,  we can easily apply Euler's formula to a probability clock. A probability clock~\cite{NOOM} is a time-varying probability distribution for a phase-space with 2 states, such that the probabilities are 
    $\cos^2(t)$ and $\sin^2(t)$, for the first and second states respectively.

A 2-dimensional real wave-function allows us to apply the Euler's formula to the probability clock:
\begin{align}
  \label{rotation}
  \Psi(t)=\exp\left(\left[\begin{smallmatrix} 0 & -1\\ 1 & 0 \end{smallmatrix}\right] t\right)\left[\begin{smallmatrix} 1\\ 0 \end{smallmatrix}\right]=\left[\begin{smallmatrix} \cos(t)\\ \sin(t) \end{smallmatrix}\right]
\end{align}

The Euler's formula for the density matrix is:
\begin{align}
  \Psi \Psi^\dagger=\left[\begin{smallmatrix} \cos^2(t) & \cos(t)\sin(t)\\ \cos(t)\sin(t) & \sin^2(t) \end{smallmatrix}\right]=\frac{1}{2}+ \left[\begin{smallmatrix} \frac{1}{2} & 0\\ 0 & -\frac{1}{2} \end{smallmatrix}\right](\cos(2t) +J\sin(2t))
\end{align}

Where $J=\left[\begin{smallmatrix} 0 & 1 \\ -1 & 0 \end{smallmatrix}\right]$ plays the role of the imaginary unit in the Euler's formula for the probability clock. A measurement using a diagonal projection triggers the collapse of the wave-function, such that a new density matrix is obtained by setting the off-diagonal part (i.e. the part proportional to $J$) of the original density matrix to zero. The probability distribution is given by the diagonal part of the density matrix, i.e. by taking the ``real part'' of the ``complex number''
$\cos(2t) +J\sin(2t)$:
\begin{align}
\label{diagonal}
  \diag(\Psi\Psi^\dagger)=\left[\begin{smallmatrix} \cos^2(t) & 0\\ 0 & \sin^2(t) \end{smallmatrix}\right]=\frac{1}{2}+\left[\begin{smallmatrix} \frac{1}{2} & 0\\ 0 & -\frac{1}{2} \end{smallmatrix}\right]\cos(2t)
\end{align}

Since $\cos^2(t)+\sin^2(t)=1$ and $0<\cos^2(t)<1$, we can confirm that the wave-function parametrizes all probability distribution functions for a phase-space with 2 states,
i.e. for any probability $p$ there is an angle $t$ such that the cosinus $\cos(t)$ of that angle verifies $\cos^2(t)=p$.
Moreover, two wave-functions are always related by a rotation
$\Psi(t+a)=\exp\left(J a\right)\Psi(t)$, for some $a$.

Note that the rotation is an invertible linear transformation that preserves the space of wave-functions. This does not happen with probability distributions: the most general linear transformation of a probability distribution that preserves the space of probability distributions is:
\begin{align}
M(a,b)=\left[\begin{smallmatrix} \cos^2(a) & \cos^2(b)\\ \sin^2(a) & \sin^2(b)\end{smallmatrix}\right]\ \ \ \ \mathrm{(where\ a,b\ are\ real\ numbers)}
\end{align}
 because if we apply $M$ to a deterministic distribution  $\left[\begin{smallmatrix} 1\\ 0 \end{smallmatrix}\right]$ or $\left[\begin{smallmatrix} 0\\ 1 \end{smallmatrix}\right]$ we must obtain probability distributions which leads to the constraints  $\cos^2(a)+\sin^2(a)=\cos^2(b)+\sin^2(b)=1$ and 
$\cos^2(a),\sin^2(a),\cos^2(b),\sin^2(b)\geq 0$; the matrix $M$ such that:
\begin{align}
M\frac{1}{2}\left[\begin{smallmatrix} 1 \\ 1 \end{smallmatrix}\right]=\left[\begin{smallmatrix} 1\\ 0 \end{smallmatrix}\right]
\end{align}
is necessarily singular and so it is not suitable to represent a symmetry group.

The wave-function is thus a good parametrization which allows us to represent a  group of transformations using linear transformations of the points of a circle. The collapse of the wave-function is nothing more than taking the real part of a complex number as in most applications of Euler's formula in Engineering, reflecting the fact that the circle is not the phase-space of the theory. Thus the wave-function is nothing more than a parametrization of the probability distribution.

\section{The Stern-Gerlach experiment}
\label{sec:sg}

We follow reference~\cite{sakurai} for the description of the Stern-Gerlach experiment, first carried out in Frankfurt by O. Stern and W. Gerlach in 1922. This experiment makes a strong case in favor of generalizing the symmetry transformations to become non-deterministic, moreover the theoretical predictions only require a phase-space with two states like the one already discussed in the previous section. Note that we only make measurements along the z and x-axis, but if we also had made measurements along the y-axis then the phase space would require four states or a parametrization with a complex wave-function, see Section~\ref{sec:complex}. Some articles such as reference~\cite{sgquantum} argue that a ``full quantum'' analysis of the Stern-Gerlach experiment must involve the position degrees of freedom and thus a phase-space with more than two states. But as in every theoretical model for any real experiment we should consider only a phase-space which is as large as it is strictly necessary to compute all predictions for all practical purposes and do not waste time with redundant calculations which only add complexity and increase the likelihood of committing mistakes. Of course, the real Stern-Gerlach experiment involves much more than two states, for instance if the electrical power feeding the experiment is shutdown due to an earthquake or if the man managing the experiment has a heart attack it will affect the experimental results, but all predictions for all practical purposes can be computed using a phase-space with only two degrees of freedom.

\begin{figure}[h!]
  \includegraphics[width=0.96\linewidth]{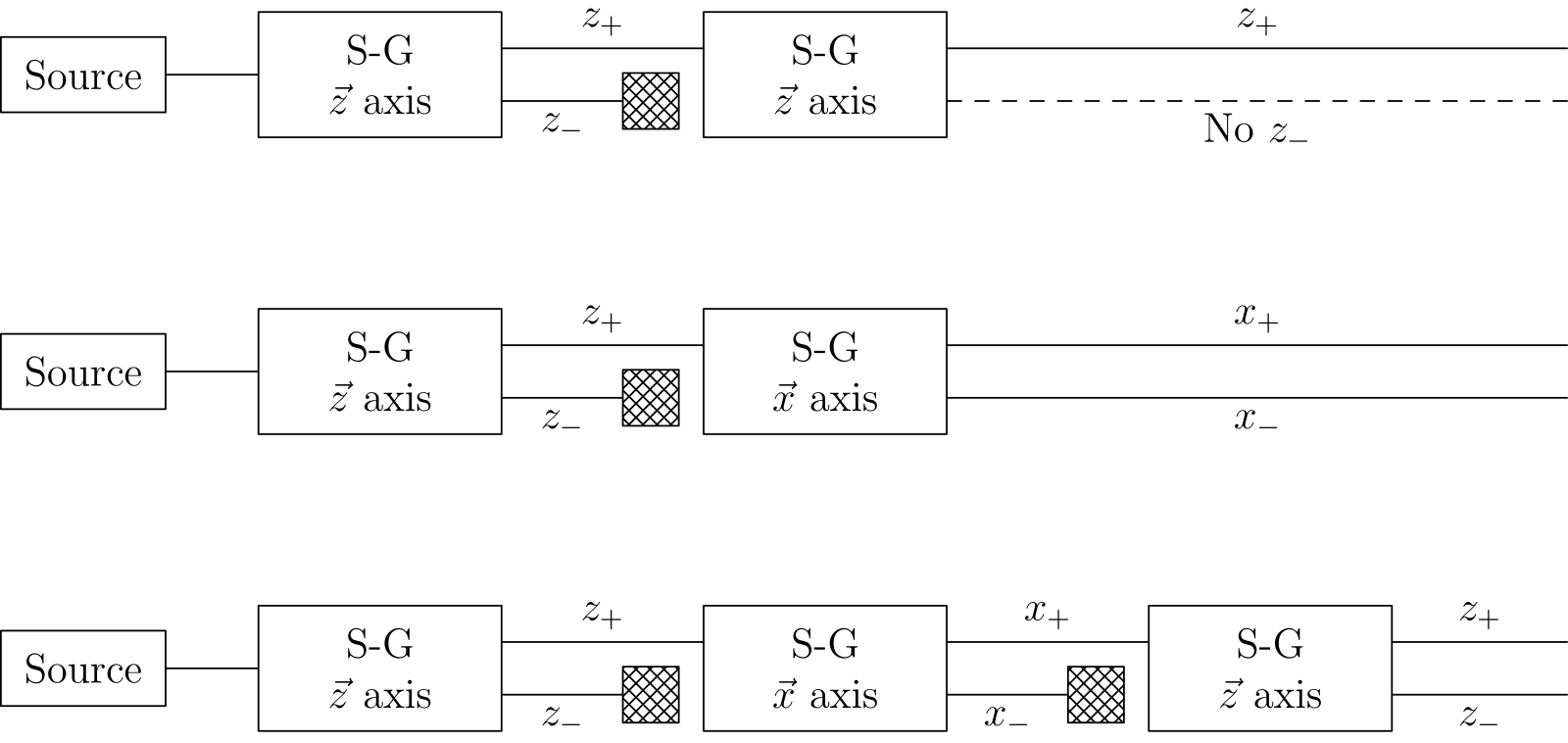}
  \caption{Sequential Stern-Gerlach experiments, figure reproduced from reference~\cite{sggraph}.}
  \label{sgseq}
\end{figure}

In the Stern-Gerlach experiment, a beam of silver atoms is sent through a magnetic field with a gradient along the z or x-axis and their deflection is observed. The results show that the silver atoms possess an intrinsic angular momentum (spin) that takes only one of two possible values (here represented by the symbols + and -). Moreover in sequential Stern-Gerlach experiments (see figure~\ref{sgseq}), the measurement of the spin along the z-axis destroys the information about a atom's spin along the x axis.

We consider in the phase-space, not only the spin of one atom of the beam, but also the angle of orientation of a macroscopic object which serves as a reference, for pedagogical purposes. The corresponding complete wavefunction is thus a reducible representation of the rotation group. When we apply a rotation to the phase-space, the rotation is a non-deterministic transformation of the spin of the atom and a deterministic transformation of the macroscopic object. Thus, to keep track of the part of the wave-function corresponding to the angle of orientation of the reference macroscopic object we only need the central value of the probability distribution for such angle, which we will call simply ``the angle'' for brevity. And then we only consider the part of the wave-function corresponding to the spin of the atom.

In Equation~\ref{diagonal}, $cos^2(t)$ is the probability for the spin to be in the state $+$, while $sin^2(t)$ is the probability for the spin to be in the state $-$. The non-deterministic symmetry transformation given by a rotation of the spin along the $x-z$ plane is parametrized by the parameter $t$ and its linear representation on the wave-function is described in Equation~\ref{rotation}.

In the first measurement, the angle of the reference macroscopic object is 0 with respect to the z-axis; and we know for sure that the spin is in the state $+$ ($t=0$) because we are measuring the spin along the z-axis of atoms that were previously filtered to be in the state $+$ when measuring the spin along the z-axis (see the first graph in figure~\ref{sgseq}). 

A second sequential measurement along the x-axis means that we rotate the reference macroscopic object 90 degrees along the x-z plane so the new angle is 90 degrees; for the atom we first make a 45-degrees rotation along the x-z plane ($t=\pi/4$)\footnote{Note that the angle is 45-degrees instead of 90-degrees because the spin group is a double cover of the orthogonal group, for which the angle would be 90-degrees.} and then we determine whether the spin is in the $+$ or $-$ state (i.e. the wave-function collapses, see  the second graph in figure~\ref{sgseq}). The probability for the spin to be in the states $+/-$ is now $50\%/50\%$, because the rotation is a non-deterministic symmetry transformation.

A third sequential measurement along the z-axis means that we rotate the reference macroscopic object -90 degrees along the x-z plane so the new angle is again 0 degrees; for the atom we first apply a -45-degrees rotation along the x-z plane ($t=-\pi/4$) to the atoms with spin $+$ and then we determine whether the spin is in the $+$ or $-$ state (i.e. the wave-function collapses one more time, see the third graph in figure~\ref{sgseq}). Despite that in the first measurement the spin was in the state $+$, the probability for the spin to be in the states $+/-$ is $50\%/50\%$ in the third measurement, because the rotation is a non-deterministic symmetry transformation and we applied it in the second and third measurements to switch from the z to the x-axis and then to switch again from the x to the z-axis.

As we have seen in the previous sections, generalizing the symmetry transformations to be non-deterministic suffices to account for all experimental results described by Quantum Mechanics, with the Stern-Gerlach experiment being one example. 
The question remaining is whether the Euler's formula applies for phase-spaces with more than 2 states, which would imply that the collapse of the wave-function is merely a mathematical artifact of the wave-function parametrization.

\section{Black hole information paradox and the Stern-Gerlach experiment}
\label{sec:informationparadox}

What is exactly a black hole from the point of view of a quantum theory?
That's a tough question. Because of that, the black hole information paradox is not necessarily related with \emph{real} black holes.

Nevertheless, we can always think of the Stern-Gerlach experiment, described in the previous section. The argument here is that there is always a unitary transformation such that the corresponding probability distribution is necessarily the constant distribution, for all initial states in the same orthogonal basis. Thus, if a black-hole erases most information about an object that comes inside of it by turning this information to random, that is not incompatible with a unitary time-evolution. We have seen an analogous case in the previous section for a 2-state phase space.

Certainly, the collapse of the wave-function is not unitary and thus the transformation on the ensemble is also not unitary. If we measure the properties of the black-hole immediately after the object comes inside, the information is erased. However since the time-evolution is unitary, if the transformation is not only about the object coming inside but about more events then the information is not necessarily lost. If such events do not affect the degrees of freedom that were erased (which is expected since a black-hole is defined by few parameters), then the information will remain erased. Only with a quantum theory for black holes we can know for sure which events can happen after an object comes inside a black-hole.

In any case, a transformation which erases information is compatible with a unitary time-evolution.

\section{Euler's formula for a phase-space with \arabic{Nstates} states}
\label{sec:mapN}

We address now a system with \arabic{Nstates} possible states. A real normalized wave-function $\varphi_1$ can be parametrized in terms of Euler angles (i.e. standard hyper-spherical coordinates and following reference~\cite{eulerangles}) as:

\begin{align}
  \WriteWaveFunction{Nstates}
%\phi=c_1 l_1 + s_1 c_2 l_2 + s_1 s_2 l_3
\end{align}

Where $c_n=\cos(\theta_n)$ and $s_n=\sin(\theta_n)$ stand for the cosine and sine of an arbitrary angle $\theta_n$ (i.e.  $\theta_n$ is an arbitrary real number), respectively; and $n$ is an integer number verifying $1\leq n<\arabic{Nstates}$. The set $\WriteBasis{Nstates}$ are normalized vectors forming an orthonormal basis of a \arabic{Nstates}-dimensional real vector space.

The Euler’s formula for the corresponding density matrices is:

\begin{align}
  \WriteDensityMatrix{Nstates}
%\phi=c_1 l_1 + s_1 c_2 l_2 + s_1 s_2 l_3
\end{align}

Where $J_n=(l_{n}\varphi_{n+1}^\dagger
     -\varphi_{n+1}l_{n}^\dagger)$ plays the role of the imaginary unit in the Euler's formula, in the subspace generated by the vectors $\{l_{n}, \varphi_{n+1}\}$.  Thus, the collapse of the wave-function for a phase-space with \arabic{Nstates} states is a recursion of collapses of 2-dimensional real wave-functions. The conditional probabilities are given by the diagonal part of the density matrix, 
    i.e. by taking the ``real part'' of the ``complex numbers'' $\cos(2\theta_n) +J_n\sin(2\theta_n)$:
\begin{align*}
  \WriteConditional{Nstates}
%  P( n | (n \mathrm{\ or\ above}))=& c_n^2\\
%  P( (n+1 \mathrm{\ or\ above}) | (n \mathrm{\ or\ above}))=& s_n^2
\end{align*}

where $P( 2 | (2 \mathrm{\ or\ above}))$ stands for probability for the state to be $n=2$ knowing that the state is either $n=2$, or $n=3$, ... or $n=\arabic{Nstates}$. Note that these conditional probabilities are arbitrary, i.e. for any probability $p$ there is an angle $\theta_n$ such that the cosinus $c_n=\cos(\theta_n)$ of that angle verifies $c_n^2=p$.
%the collapse of the wave-function of any quantum system is a recursion of collapses of 2-dimensional real wave-functions.

The fact that the previous conditional probabilities are arbitrary, implies that the probability distribution is arbitrary, since for any probability distribution we have:
\begin{align}
\WriteBornConditional{Nstates}
\end{align}
%The Born rule gives for the probabilities of the events corresponding to vectors $\WriteBasis{Nstates}$:
Moreover, two wave-functions are always related by a rotation. Thus we can confirm that any probability distribution for \arabic{Nstates} states, can be reproduced by the Born rule for some wave-function:
\begin{align}
P(n)=&|\varphi^\dagger l_n|^2\\
\WriteBorn{Nstates}
\end{align}

\section{Euler's formula for a generic phase-space}
\label{sec:map}

A probability distribution can be discrete or continuous. A continuous probability distribution is a probability distribution that has a cumulative distribution function that is continuous. Thus, any partition of the phase-space (where each part of the phase-space has a non-null Lebesgue measure) is countable.

Consider now a countable (possibly infinite) partition of the phase-space. The corresponding countable orthonormal basis for the separable Hilbert space is $\{l_n\}$, where each index $n>0$ corresponds to an element of the partition of the phase-space.
We can parametrize a normalized vector in the Hilbert space~\cite{eulerangles}, as $v_n=c_n l_n+s_n v_{n+1}$, where $c_n=\cos(\theta_n)$ and $s_n=\sin(\theta_n)$ stand for the cosine and sine of an arbitrary angle $\theta_n$ (i.e.  $\theta_n$ is an arbitrary real number), respectively; and $n>0$ is an integer number. The first vector $v_1$ is the wave-function of the full phase-space. Note that the parametrization is valid for infinite dimensions, because in the recursive equation all we need to assume about the vector $v_{n+1}$ is that it is normalized and orthogonal to $\{l_1, l_2, ... l_n\}$, which is a valid assumption in infinite dimensions. Then we define $v_{n+1}$ in terms of $v_{n+2}$ in the same way, and so on. The recursion does not need to stop.

Then, the projection to the linear space generated by $v_n$ is:
\begin{align}
  v_nv_n^\dagger=&\frac{1}{2}+\frac{1}{2}(l_{n}l_{n}^\dagger-\varphi_{n+1}\varphi_{n+1}^\dagger) (\cos(2\theta_{n})+J_{n}\sin(2\theta_{n}))
 \end{align} 

Where $J_n=(l_{n}\varphi_{n+1}^\dagger-\varphi_{n+1}l_{n}^\dagger)$ plays the role of the imaginary unit in the Euler's formula, in the subspace generated by the vectors $\{l_{n}, \varphi_{n+1}\}$.  Thus, the collapse of the wave-function for a generic phase-space is a recursion of collapses of 2-dimensional real wave-functions. 
The conditional probabilities are given by the diagonal part of the density matrix, i.e. by taking the ``real part'' of the ``complex numbers'' $\cos(2\theta_n) +J_n\sin(2\theta_n)$:
The operator $v_n v_n^\dagger$ is a projection thanks to the off-diagonal\footnote{In a basis where all $l_n l_n^\dagger$ are diagonal.} terms $c_ns_n(l_n v_{n+1}^\dagger+v_n l_{n+1}^\dagger)$.

Defining $(n\mathrm{\ or\ above})=\{k : k\geq n\}$ as the event which contains all parts of the phase-space with index starting at $n$, we can write the probability distribution as:
\begin{align}
\label{eq:cond}
  P(n)&=P((n\mathrm{\ or\ above}))P(n| (n \mathrm{\ or\ above}))\\
  &=\left(\prod\limits_{k=1}^{n-1} P((k+1 \mathrm{\ or\ above})|(k \mathrm{\ or\ above}))\right)P(n| (n \mathrm{\ or\ above}))
\end{align}

That is, as a product of the probabilities 
\begin{align}&P(n|(n \mathrm{\ or\ above}))\mathrm{\ and\ }P((n+1 \mathrm{\ or\ above})|(n \mathrm{\ or\ above}))\mathrm{,\ which\ verify}\\
&P(n|(n \mathrm{\ or\ above}))\ +\ \ 
P((n+1 \mathrm{\ or\ above})|(n \mathrm{\ or\ above}))=1.
\end{align}

If the off-diagonal terms are suppressed (collapsed), we obtain a diagonal operator which represents the probability distribution $P(n)$ in the Hilbert space:

\begin{align}
  \diag(v_nv_n^\dagger)=c_n^2 l_nl_n^\dagger+s_n^2 v_{n+1}v_{n+1}^\dagger
\end{align}

That is, $P(n)=\tr(\diag(v_1v_1^\dagger) l_nl_n^\dagger)$ and $P(O)=0$ for operators $O$ with null-diagonal.
Note that $c_n^2=P(n| (n \mathrm{\ or\ above}))$ and $s_n^2=P((n+1 \mathrm{\ or\ above})| (n \mathrm{\ or\ above}))$ and these probabilities are arbitrary, i.e. for any probability $p$ there is an angle $\theta_n$ such that the cosinus $c_n=\cos(\theta_n)$ of that angle verifies $c_n^2=p$. 

The fact that these conditional probabilities are arbitrary, implies that the probability distribution is arbitrary, since the probability distribution can be written in terms of these conditional probabilities as shown in Equation~\ref{eq:cond}.

\section{Complex and Quaternionic Hilbert spaces}
\label{sec:complex}

While the parametrization with a real wave-function is always possible, it may not be the best one. As we have seen, the wave-function parametrization allows us to apply group theory to the states of the ensemble, since  unitary transformations (i.e. a multi-dimensional rotation) preserve the properties of the parametrization (in particular the conservation of total probability).

The union of a set of projection operators and the unitary representation of a group, is a set of normal operators. Suppose that there is no non-trivial closed subspace of the Hilbert space left invariant by this set of normal operators. The (real version of the)  Schur's lemma~\cite{realpoincare,Oppio:2016pbf,realoperatoralgebras} implies that  the set of operators commuting with the normal operators forms a real associative division algebra---such division algebra is isomorphic to either: the real numbers, the complex numbers or the quaternions.

If we do a parametrization by a real wave-function and consider only expectation values of operators  that commute with a set of operators  isomorphic to the complex or the quaternionic numbers, then we can equivalently define wave-functions in complex and quaternionic Hilbert spaces~\cite{realQM,realpoincare,Oppio:2016pbf}.

%Note that the cases we are interested in are not the most general possibilities. Certainly, real numbers can be considered as a particular case of complex numbers and these in turn can be considered a particular case of quaternionic numbers. However, we are working with multi-dimensional Hilbert spaces in the context of group theory. The real irreducible representations are always more general than the corresponding complex or quaternionic ones, because they allow more transformations to be performed on them\footnote{Already the fact that complex and quaternionic numbers are isomorphic to \emph{real} division algebras, is a hint that is the case.}. For instance, the real fundamental representation of the gauge group $SU(2)_L$ allows to define custodial transformations on them, this has important phenomenological consequences in the Standard Model.

Let us consider the quaternionic case (it will be then easy to see how is the complex case). We have a discrete state space defined by two real numbers $n,m$, with $1\leq m\leq 4$ and we only consider the probabilities for $n$ independently on $m$, $P(n)=\sum_{m=1}^4 P(n,m)$.

Then a more meaningful parametrization---reflecting by construction the restriction on the operators we are considering---uses a quaternionic wave function $v_1$. Let $\{l_n\}$ be an orthonormal basis of quaternionic  wave-functions and we have:
\begin{align}v_n v_n^\dagger=c_n^2 l_nl_n^\dagger+s_n^2 v_{n+1}v_{n+1}^\dagger+c_ns_n(l_n v_{n+1}^\dagger+v_{n+1} l_{n}^\dagger)
\end{align}

Note that there is a basis where $l_nl_n^\dagger$ is real diagonal and thus upon collapse $v_nv_n^\dagger$ becomes real diagonal as well.

The complex case is just the above case with complex numbers replacing quaternions and a state space which is the union of 2 identical spaces. The continuous case is analogous, since there is a partition of the phase-space which is countable.

\section{Comparing the time evolution with a stochastic process}
\label{sec:stochastic}

Quantum Mechanics is not a generalization of probability theory, but
it is definitely a generalization of classical mechanics since it
involves non-deterministic transformations to the state of the
system. For instance, the time evolution may be non-deterministic unlike in classical mechanics.

There are three major metaphysical views of time~\cite{bebecome}: presentism, eternalism and possibilism. The possibilism consists in considering the presentism for the future and the eternalism for the past, so it is inconsistent with a time translation symmetry.
The presentism view coincides with the Hamiltonian formalism of physics, that the state of the system is defined by a point in the phase space.
When the time evolution of the system is deterministic it traces a
phase space trajectory for the system, however the definition of the
state of the system does not involve time, i.e. only the present
exists~\footnote{The presentism view is compatible with
  (non-quantum) special or general
  relativity\cite{diracdynamics,adm}. There are difficulties with the
  quantum versions of relativity (more with general relativity than
  with special relativity~\cite{wigner}), but these difficulties have little to do with the presentism as we will discuss in forthcoming articles.}.
The eternalism view coincides with the Lagrangian formalism of
physics, that the state of the system is defined by a function of
time. When the time evolution of the system is deterministic, this
function of time coincides with the phase-space trajectory of the
classical Hamiltonian formalism and so which metaphysical view of time we use is
irrelevant from an experimental point of view (in the deterministic case). 

But when the time-evolution of the system is non-deterministic, we may have a hard time studying the
time-evolution from the Lagrangian formalism and/or eternalism metaphysical view.
The key fact about Quantum Mechanics which makes it incompatible with the eternalism/Lagrangian point of view is that the time-evolution is not necessarily a stochastic process, i.e. there is not necessarily a collection of random events indexed by time\footnote{It is in possible to insist that the time-evolution is always a stochastic process, but then we must consider a different theory than Quantum Mechanics, e.g. described by the Lindblad equation instead of the Schrödinger equation~\cite{WeinbergLindblad}.}.
We only apply one non-deterministic transformation of the state of the
system, however there are many different transformations we can choose
from and the set of choices is indexed by a parameter we call time, which is fine from the presentism/Hamiltonian point of view since only the present exists.

Note that a random  experiment always involves a
preparation followed by a measurement. For instance, we shake a dice
in our hand and throw it over a table until it stops (preparation),
then we check the position where it stopped (measurement).

If we just throw the dice without shaking our hand, the
probability distribution for the measurement outcome is different than
if we shake our hand. There is nothing mysterious about this:
two different preparations lead to two different probability distributions.
%;this is effectively the case of two different theories leading to two different predictions. 
Whether or not we actually do the measurement does not change
anything, what changes the probability distribution is the
preparation.

Then we can think about a preparation which is function of an element
of a symmetry group, for instance translation in time. From the point
of view of probability theory or experimental physics, this is a valid
option.
However, it is important to note that this preparation function of
time is not a stochastic process in time.
A stochastic process in time is a set of random experiments indexed by
time, while in the preparation which is function of time we have a
single random experiment dependent on the parameter time.
As an example, consider a) throwing the dice 10 times,
one time per minute during 10 minutes
and b) shake the dice in our hand for a number of minutes $T$ between 0
and 10 and then throw the dice once.
The preparation in b) is dependent on the time parameter $T$, while in a)
the time selects the one of the many identically prepared experiments
which was done at the selected time.

Note that the experiments a) and b) above are different but can be
combined: we could do many random experiments, each of them would be
dependent on a parameter. This fact is important in Section~\ref{sec:timetranslation}.

In the remaining of this section, we comment on conditioned probability and the random walk.
It is well-known that quantum mechanics can be described as the Wick-rotation of a Wiener stochastic process~\cite{nonperturbativefoundations}. In other words, the time evolution in Quantum Mechanics is a Wiener process for imaginary time. This is the origin of the Feynman's path integral approach to Quantum Mechanics and Quantum Field Theory.

Since the Wiener process is one of the best known L\'{e}vi processes---a L\'{e}vi process is the continuous-time analog of a random walk---this fact often leads to an identification of Quantum Mechanics with a random walk. In particular, it often leads to an identification of the probabilities calculated in Quantum Mechanics with conditioned probabilities---the next state in a random walk is conditioned by the previous state.

Certainly, the usefulness of group theory is common to both a random walk and to Quantum Mechanics and this unavoidably leads  to similarities between  a random walk and  Quantum Mechanics.
However, imaginary time is very different from real time and thus the probabilities calculated in Quantum Mechanics are not necessarily conditioned probabilities in a random walk.

In order to relate a random walk (or any other stochastic process) with Quantum Mechanics correctly, we need the probability distribution for the complete paths of the random walk. Then, we can use a wave-function parametrization of the probability distribution for the complete paths of the random walk. Finally, we can apply quantum methods to this wave-function. The result is a Quantum Stochastic Process~\cite{qsc}, which is not a generalization of a stochastic process due to the wave-function collapse, but merely the parametrization of a stochastic process with a wave-function.

\section{Time translation is a
stochastic process if and only if it is deterministic}
\label{sec:timetranslation}

Now we are able to prove one of the main results of this paper, namely that there is a group action of a Wigner's symmetry group on the probability distribution for the state of a system, if and only if the Wigner's symmetry group
transforms deterministic (probability) distributions into
deterministic (probability) distributions. A corollary is that time
translation in Quantum Mechanics is a stochastic process if and only if it is deterministic.  This mathematical fact is overlooked by the assumptions of both the Bell's theorem and the Einstein-Podolsky-Rosen (EPR) paradox.

As it was discussed in Section~\ref{sec:symmetries}, Wigner's theorem~\cite{2014PhLA..378.2054G,Ratz1996,*wignertheorem} implies that the action of a symmetry group on the wave-function is necessarily linear and unitary.
In Section~\ref{sec:deterministic}, we showed that the action of a symmetry
group on the wave-function is deterministic if and only if $P_A$ and $U_g P_B U^\dagger_g$
commute for all events $A,B$ and for all the elements $g$ of the group,
where $P_A$ is a projection-valued-measure.

This means that $U$ is a deterministic transformation if and only if $U_{l a} U_{m a}^*=0$
for all $a,l,m$ such that $l\neq m$.

Now we check the necessary and sufficient conditions for the action of
a symmetry group on the wave-function to correspond to an action on
the corresponding probability distribution. 

That is, if we start with some probability distribution $\diag(\rho_1)$,
then the action of each element $g$ of the group on the wave-function will produce 
(after the collapse) a different probability distribution $\diag(\rho_g)$.
The composition of the actions of two group elements $g,h$ on the
probability distribution is given by the succession of the two random experiments
corresponding to $g$ and $h$: $P(A)=\tr(\diag(\rho_g)U_h P_A U_h^\dagger)$.

However, Wigner's theorem~\cite{2014PhLA..378.2054G,Ratz1996,*wignertheorem} implies
that the action of a symmetry group on the wave-function is necessarily linear and unitary,
thus $P(A)=\tr(\rho_g U_h P_A U_h^\dagger)$.

Thus there is a group action of the symmetry group on the probability
distribution if and only if $\tr(\diag(\rho_g)U_h P_A U_h^\dagger)=\tr(\rho_g U_h P_A U_h^\dagger)$ 
for any pure density matrix $\rho_g$ and any event $A$ and group
element $h$.

The equality above is equivalent to $\sum_{k,b: k\neq b}
U_{ka}^*\Psi_k\Psi_b^* U_{ba}=0$, where $U_{ba}$ are the elements of the
matrix $U_h$. We can see
that if $U$ is a deterministic transformation, then the equality is
satisfied, since $U_{ja}^* U_{jl}=0$ for all $a,l,j$ such that $a\neq
l$. On the other hand, if $U$ is a non-deterministic transformation
then for some $a,l,m$ such that $l\neq
m$, we have $U_{m a}^* U_{l a}\neq 0$.
Then for $\Psi_k=\frac{1}{\sqrt{2}}(\delta_{km}+\delta_{kl})$, we get
$\sum_{k,b: k\neq b} U_{ka}^*\Psi_k\Psi_b^* U_{ba}=U_{m a}^*
U_{la}\neq 0$, i.e. there is no group action of the symmetry group on the probability
distribution.

\section{Symmetries as irreversible processes}
\label{sec:irreversible}

The concept of (ir)reversible process from thermodynamics also needs a careful discussion in quantum mechanics.
A non-deterministic symmetry transformation, when acting on a deterministic ensemble increases the entropy of the ensemble after the wave-function collapse and therefore must be an irreversible transformation. Yet, a symmetry transformation always has an inverse symmetry transformation, because it is included in a symmetry group, so it must be considered reversible in some sense. 

The way out of this apparent contradiction is the role of time in the quantum formalism, which was discussed in Sections~\ref{sec:stochastic} and~\ref{sec:timetranslation}. In the ensemble interpretation, the individual system is entirely defined by a standard phase-space, which implies that the time plays no fundamental role in quantum mechanics nor in classical Hamiltonian mechanics. Then, time-evolution in quantum mechanics is not a stochastic process unless it is deterministic. Therefore, there is not a probability distribution for each time (or for other parameter corresponding to the symmetry group). 

If we consider a stochastic process with only two probability distributions corresponding to the initial and final times, then the complete symmetry transformation is irreversible (if it is non-deterministic and it acts in a deterministic ensemble). However, this does not imply that it is a ``bad'' symmetry, because no stochastic process can be defined in between the initial and final times. On the other hand, if the symmetry group contains only deterministic transformations then a stochastic process can be defined in between the initial and final times and such process is reversible, as expected.

\section{Quantum Mechanics is EPR-complete}
\label{sec:complete}

The Einstein-Podolsky-Rosen (EPR) main claim~\cite{epr} (namely, that Quantum Mechanics is an incomplete description of physical reality), is defended by reducing to absurd the negation of the main claim, i.e. by reducing to absurd that position (Q) and momentum (P) are not simultaneous elements of reality. In the EPR article it is stated: ``\emph{one would not arrive at our conclusion if one insisted that two or more physical quantities can be regarded as simultaneous
elements of reality only when they can be simultaneously measured or predicted.\emph{[...]} This makes the reality of P and Q depend upon the process of measurement carried out on the first system, which does not disturb the second system in any way. No reasonable definition of reality could be expected to permit this.}''

The reduction to absurd of the negation of the claim, could only be a
satisfactory argument if the claim itself (namely, the quantities
position and momentum of the same particle are simultaneous elements
of reality, despite they cannot be simultaneously
measured or predicted)
would not be absurd as well. But the claim itself raises eyebrows to
say the least, once we remember that (in Quantum Mechanics, by
definition) measuring the position with infinite precision completely
erases any knowledge about the momentum of the same
particle.

In Quantum Mechanics as in classical Hamiltonian
mechanics, the state of an individual system is a point in a phase
space, and the phase space is both the domain and image of the
deterministic physical transformations. As in any statistical theory,
we may know only the probability distribution for the state of the
individual system, instead of knowing the state of the individual system.
The relation between quantum mechanics and a statistical theory is clear: the wave-function is a
parametrization for any probability distribution~\cite{parametrization}. 

There are two kinds of incompleteness in a non-Markov stochastic
process. The two kinds of incompleteness are in correspondence with the two
concepts: stochastic and non-Markov, respectively.
 
1) Stochastic: From the point of view of (classical) information theory~\cite{info},
the root of probabilities (i.e. non-determinism) is by definition the absence of
information. Statistical methods are required whenever we lack
complete information about a system, as so often occurs when the
system is complex~\cite{bertinstatistical}. Thus we can convert a
deterministic process to a stochastic process unambiguously (using
trivial probability distributions); but we cannot convert a
stochastic process into a deterministic process unambiguously since we
need new information~\footnote{E.g. the assumptions required by the
deterministic models in reference~\cite{automaton} are new
information.}.

2) non-Markov: any non-Markov stochastic process can be described as a
Markov stochastic process where some variables defining the state
of the system are hidden (i.e. unknown)~\cite{allmarkov,non_markov_examples}.
Conversely, by definition any irreducible~\footnote{The
word irreducible avoids the case where the state of the system is
the direct product of two states corresponding to 2 
irreducible Markov processes, in such case the fact that some
variables are hidden does not imply that the resulting stochastic
process is non-Markov.} Markov process where some variables defining the state
of the system are hidden will give rise to a non-Markov process. For
instance, the physical phenomena which generates examples of brownian
motion is deterministic and thus Markov, but real-world
brownian motion is often non-Markov (because we cannot measure the
state of the system completely~\cite{brownian,brownian2}) despite
the fact that the brownian motion is one of the most famous examples
of a Markov process.

In reference~\cite{reality} (authored by A. Einstein and contemporary
of the EPR paradox) the two kinds of incompleteness are clearly
distinguished:

``\emph{\emph{[...]} I believe
that the \emph{[quantum]} theory is apt to beguile us into error in our search for
a uniform basis for physics, because, in my belief, it is an
incomplete representation of real things, although it is the
only one which can be built out of the fundamental concepts
of force and material points (quantum corrections to classical
mechanics). The incompleteness of the representation is the
outcome of the statistical nature (incompleteness) of the laws.
I will now justify this opinion.}''

The incompleteness of the representation corresponds to the non-Markov
kind, while the incompleteness of the laws corresponds to the stochastic
kind. By definition, in Quantum Mechanics any sequence of measurements is a Markov stochastic
process (thus it has the stochastic kind of incompleteness)~\footnote{The term (non-)Markov stochastic process is used in this paper in
the classical sense as in reference~\cite{non_markov}. We do not mean a quantum (non-)Markov
stochastic process as in reference~\cite{non_markov}, despite the fact that Quantum Mechanics
is used to calculate the transition probabilities between measurements.}. Note that any
non-Markov stochastic process can be described as a Markov stochastic process
where some variables defining the state of the system are hidden
(i.e. unknown)~\cite{allmarkov,non_markov_examples}.

Since Quantum Mechanics does not have the non-Markov kind of incompleteness, position and momentum can only be simultaneous elements of reality in another theory very different from Quantum Mechanics. That
both the claim and its negation are absurd, is strong evidence that
some of the assumptions leading to the Einstein-Podolsky-Rosen (EPR)
paradox~\cite{epr} do not hold.

So, why did the author tried to justify (using the EPR
paradox~\cite{epr}, among other arguments) that in Quantum Mechanics
the stochastic kind of incompleteness necessarily leads to a
non-Markov kind of incompleteness?

The following paragraph from the same reference~\cite{reality} suggests
that the author was trying to favor the cause that any future
theoretical basis should be deterministic, not just Markov (since
statistical mechanics is often Markov). 

``\emph{There is no doubt that quantum mechanics has seized hold
of a beautiful element of truth, and that it will be a test stone
for any future theoretical basis, in that it must be deducible
as a limiting case from that basis, just as electrostatics is
deducible from the Maxwell equations of the electromagnetic
field or as thermodynamics is deducible from classical mechanics.
However, I do not believe that quantum mechanics
will be the starting point in the search for this basis, just as,
vice versa, one could not go from thermodynamics (resp.
statistical mechanics) to the foundations of
mechanics.}''

However and as discussed in Section~\ref{sec:deterministic}, there is no
mathematical argument that suggests that in general a deterministic model is more
fundamental than a stochastic one, quite the opposite. Since the wave-function is merely a possible parametrization of any
probability distribution~\cite{parametrization}, we also cannot claim that a  deterministic model is more
fundamental than Quantum Mechanics. Thus, the stochastic kind of
incompleteness is harmless. 

So, the EPR paradox appears as an attempt to justify a mathematical
statement (that a  deterministic model is more
fundamental than Quantum Mechanics) with arguments
from physics (trying to link to the non-Markov kind of incompleteness), for which no mathematical
arguments could be found. Note that a statement referring to
any future theoretical basis is essentially a mathematical statement because the
physical model is any (since the theoretical basis is any).

However, it is a failed attempt because it missed the fact discussed in
Section~\ref{sec:timetranslation}, that the time evolution is a stochastic process
if and only if it is deterministic.

In the EPR paradox, there is no probability distribution for the state
of system after the spatial separation of the entangled particles and before the
transformation involved in the measurement takes place, because the time evolution (being in this
case non-deterministic) is not a stochastic process.
We can only consider the probability distribution for the state
of system after the spatial separation of the entangled particles and after the
transformation involved in the measurement takes place. This is overall a non-local physical
transformation since it involves the spatial separation of the
entangled particles. But it does not violate relativistic causality, since both the spatial separation of the
entangled particles and the transformation involved in the measurement do not by themselves
violate relativistic causality, so their composition does not violate
causality either.

Unlike many popular no-go arguments~\cite{nogo}, we are not arguing
against the requirement that a physical theory should be complete, in
fact we claim that Quantum Mechanics is a complete statistical
theory (as defined by EPR). 

Note that Bohr already declared Quantum Mechanics as a ``complete''
theory, however he did it at the cost of a radical revision of the classical notions of
causality and physical reality~\cite{bohrcomplete}. He wrote: 
``Indeed the finite interaction between
object and measuring agencies conditioned by the very existence of the quantum of action entails
---because of the impossibility of controlling the
reaction of the object on the measuring instruments
if these are to serve their purpose---the
necessity of a final renunciation of the classical
ideal of causality and a radical revision of our
attitude towards the problem of physical reality.''~\cite{bohrcomplete}
Such notion of a ``complete'' theory mostly favours the EPR claim:
the only way that Quantum Mechanics could be complete is if it is
incompatible with the classical notions of
causality and physical reality. Thus from a logic point of view,
there is no disagreement between Einstein and Bohr, their disagreement
is about what basic features an acceptable theory should have, whether or not
it should be compatible with the classical notions of
causality and physical reality.

In contrast, the fact---that the time evolution is a stochastic process
if and only if it is deterministic---which was overlooked is perfectly
compatible with the classical notions of physical
reality (because Quantum Mechanics has a standard phase-space) and causality 
(as we will show in Section~\ref{sec:relativistictheory}). We claim that Quantum
Mechanics---being non-deterministic and thus a generalization of
classical mechanics---does not entail a radical departure from the
basic features that an acceptable theory should have, according to
EPR~\cite{epr}. In fact in Quantum Mechanics and in classical Hamiltonian
mechanics, the state of an individual system is a point in a phase
space, and the phase space is both the domain and image of the
deterministic physical transformations.

\section{Any deterministic theory compatible with
  relativistic Quantum Mechanics necessarily respects relativistic
  causality}
\label{sec:relativistictheory}

The only known theory consistent with the experimental results in high energy physics~\cite{pdg}
is a quantum gauge field theory which is mathematically
ill-defined~\cite{prize}. Due to the mathematically illness, the
relation of such a theory with Quantum Mechanics is still object of
debate and it will be addressed soon in another article by the present
author.

In the mean time we will have to consider a free system, which
suffices to address the EPR paradox.
For a free system, we know well what is relativistic Quantum
Mechanics~\cite{realpoincare}. The time evolution of the wave-function
is described by the Dirac equation for a free particle, which is a real (i.e. non-complex)
equation.

Relativistic causality is satisfied in relativistic Quantum Mechanics, meaning that there is a
propagator which vanishes for a space-like propagation~\cite{realpoincare}. In other words,
the probability that the system moves faster than light is null.

A deterministic theory compatible with
relativistic Quantum Mechanics is one which when applied to an
ensemble of free systems,  will reproduce the
statistical predictions of Quantum Mechanics.

Since in relativistic Quantum Mechanics the probability that the
system moves faster than light is null,
then no system (described by the deterministic theory) in the ensemble
moves faster than light. Thus any deterministic theory compatible with
  relativistic Quantum Mechanics necessarily respects relativistic
  causality. The question we left open here and address in the next
  section, is whether one such deterministic theory exists.

\section{A  deterministic theory compatible with relativistic Quantum
Mechanics}
\label{sec:deterministictheory}

Does a deterministic theory---consistent
with the non-deterministic time evolution of Quantum Mechanics---exists? 

The answer is yes, and we will build one example of such deterministic
theory in this section.

In an experimental setting, we always have a discrete set of possible
outcomes and thus Quantum Mechanics always predicts a cumulative
distribution function. This allows us to apply the inverse-transform
sampling method~\cite{sampling} for generating pseudo-random numbers
consistently with the probability distribution predicted by Quantum
Mechanics.

An experiment in Quantum mechanics always involves the repetition of
an experimental procedure many times. In the deterministic theory
however, each time we execute the experimental procedure we are not
executing exactly the same experimental procedure. We consider a
number (any number will do) which will be the seed of the
pseudo-random number generator and then we generate pseudo-random numbers
consistently with the probability distribution predicted by Quantum
Mechanics. The experimental procedure is: 1) generate one pseudo-random
number and 2) modify the state of
the system accordingly with the pseudo-random number.

In the case of relativistic Quantum Mechanics, the probability of
violating relativistic causality is null. Thus, the experimental
procedure never violates relativistic causality. The modifications of
the state of the system are however necessarily not infinitesimal
since the phase space of the experimental setting is discrete. This
doe not violate relativistic causality, since the finite modifications
to the state of the system occur in finite intervals of time.

We can however consider intervals of time as small as we like and thus
modifications to the state of the system as small as we like. The only
requirement for this is that the
computational resources involved in the pseudo-random number
generation are as large as needed (which is valid from a logical point
of view). Note that since time evolution in quantum mechanics is not necessarily a stochastic
process, we will often have that a sequence of experimental procedures
executed at regular and small intervals of time
produces different statistical data than than just one
experimental procedure executed at once after the same total time has
passed (e.g. in the double-slit experiment). But this cannot be considered a radical departure of the
classical notion of physical reality, since in the (very old) presentism view of
classical Hamiltonian mechanics, the phase space (i.e. the physical
reality) does not involve the notion of
time~\cite{bebecome}. Moreover when the time evolution is deterministic
then it is a stochastic process, therefore if we study only
deterministic transformations then we can recover the eternalism view of
classical Lagrangian mechanics without any conflict with relativistic
causality. For instance, this implies that in the double-slit
experiment we can in principle reconstruct the trajectory of each
particle and conclude about which slit the particle has went through.

From a logical point of view, this deterministic theory is valid and
by definition it always agrees with the experimental predictions of
Quantum Mechanics, thus it is experimentally indistinguishable from
Quantum Mechanics.

From the metaphysics point of view, this deterministic
theory is unacceptable, since it involves pseudo-random number
generation. For instance, in the double-slit experiment we (or some
super-natural entity) would need
to somehow ``program'' each particle to follow a different path determined by a
different number, which is absurd. However, the present author has no interest in
building a nice deterministic theory compatible with Quantum
Mechanics, for the reasons exposed in Section~\ref{sec:deterministic}. 

Note that this deterministic theory is not super-deterministic,
i.e. the experimental physicists are free to choose which measurements
and which transformations of the state of the system to
do~\cite{superdeterminism}. However, an
experimental procedure involves a symmetry transformation of the state of the
system. Since the symmetry transformation in this deterministic
theory is reproduced by the pseudo-random number generation, then when
we apply the inverse-transform
sampling method we need to know already what is the symmetry
transformation. Thus there is a kind of conspiracy between the symmetry
transformation and the pseudo-random generator, but such conspiracy
is part of the definition of the deterministic symmetry transformation itself.
There are assumptions about freedom of choice
in the literature which exclude our deterministic (but not
super-deterministic) theory, because the
authors erroneously consider that an experimental procedure which involves a transformation of the
state of the system is instead an observation without consequences to the system~\cite{superdeterminism}.

\section{The Young's double slit experiment}
\label{sec:doubleslit}

The ensemble interpretation does not give any explanation as to why it looks like the electron's wave-function interferes with itself in the Young's double-slit experiment~\cite{c2slit,feynmandoubleslit,bookyoung}---that would imply that the wave-function describes (in some sense) an individual system. We will fill that gap in this section.

\begin{figure}[h!]
    \includegraphics[width=0.49\linewidth]{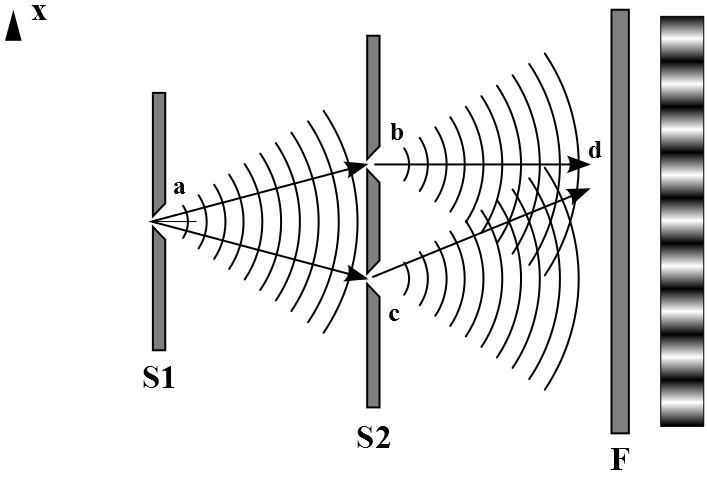}
    \includegraphics[width=0.49\linewidth]{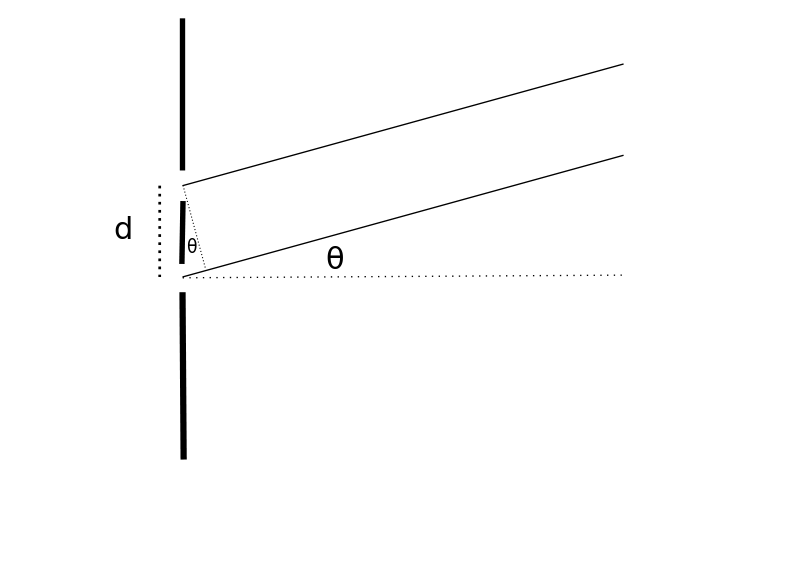}
    \caption{Double slit experiment, pictures as in references~\cite{doubleslit,doubleslit2}. In short, an electron beam is fired (S1) against a wall with two slits in it (S2). A movable mask may block the electron beam, only allowing the electrons to go through one or another slit, or instead allowing the electrons to go through both slits. Then the electrons reach the backstop and detector (F), where they are detected one by one in a corresponding spacial point. The probability distribution for the electrons that pass through the double-slit is shown next to (F). In the far-field approximation, where the distance between the slits is much smaller than the distance from the doubleslit until the detector, the difference in size between the two possible straight paths from the double slit until the detector is $d \sin(\theta)$, as illustrated in the picture on the right.}
    \label{doubleslit}
\end{figure}

The key to understand the results of the double-slit experiment is the role of time in the quantum formalism, which was discussed in detail in Section~\ref{sec:timetranslation}. In the ensemble interpretation the individual system is entirely defined by a standard phase-space, which implies that the time plays no fundamental role in quantum mechanics nor in classical Hamiltonian mechanics. Moreover, the time-evolution in quantum mechanics is not a stochastic process unless it is deterministic. Therefore, there is not a probability distribution for each time (or for other parameter corresponding to the symmetry group).

In the double-slit experiment, the time-evolution of the electron after being fired (S1) is a product of two non-deterministic symmetry transformations: first, going through one or another slit with a 50/50 probability (S2); and second, a non-deterministic propagation from (S2) until (F). If at least one of these two symmetry transformations would be deterministic, then we could define a stochastic process including the 3 instants in time (S1), (S2) and (F). But since both transformations are nondeterministic, the only stochastic process that can be defined only includes the 2 instants in time (S1) and (F), and the corresponding transformations from (S1) to (S2) and from (S2) to (F) have never occurred.

The only ``mistery'' that needs to be clarified is the fact that the non-deterministic propagation of the electron from (S2) until (F) is such that it appears that the electron interferes with itself, just like a classical wave would do. To simplify the discussion we will only consider the electrons that reach the detector along 2 different angles $\sin(\theta_1)=\frac{2 \pi}{p d}$ and $\sin(\theta_2)=\frac{\pi}{p d}$ where $p$ is the electron's linear momentum. So, a selected electron can only go through one of these 2 angles, the electrons that go through other angles are discarded.
 
The wave-function at (S1) is $\Psi=\left[\begin{smallmatrix} 1 \\ 0\end{smallmatrix}\right]$.
The time-evolution from (S1) until (S2) may be the identity matrix 
$U=\left[\begin{smallmatrix} 1 & 0 \\ 0 & 1 \end{smallmatrix}\right]$ or 
$U=\frac{1}{\sqrt{2}}\left[\begin{smallmatrix} 1 & 1 \\ 1 & -1 \end{smallmatrix}\right]$, 
  depending on whether the second slit is closed or open, respectively. If the second slit is open, then $U\Psi=\frac{1}{\sqrt{2}}\left[\begin{smallmatrix} 1 \\ 1 \end{smallmatrix}\right]$ meaning that the electron may go through both slits with equal probability.

The time-evolution from (S2) until (F) is given by the unitary transformation 
$U^{'}=\frac{1}{\sqrt{2}}\left[\begin{smallmatrix} 1 & 1 \\ 1 & -1 \end{smallmatrix}\right]$, 
that is, it sums the wave-functions from both slits for the first angle and it subtracts the wave-functions from both slits for the second angle.

Thus, if the second slit is closed, we have at (F) the wave-function 
$\Psi=\frac{1}{\sqrt{2}}\left[\begin{smallmatrix} 1 \\ 1 \end{smallmatrix}\right]$ meaning that the electron may come along angles 1 or 2 with equal probability\footnote{Strictly speaking, the larger the angle the less probability, but this fact can be compensated by collecting the electrons coming along the larger angle within a larger angular width, such that the total probabilities for angles 1 and 2 are normalized to be equal.}.
But if the second slit is open, we have at (F) the wave-function 
$\Psi=\left[\begin{smallmatrix} 1 \\ 0\end{smallmatrix}\right]$ meaning that the electron will only come along angle 1; since the electron would have come through both slits with equal probability if we would see what happened at (S2), it appears that from (S2) until (F) it interferes with itself constructively(destructively) along the angle 1(2) respectively.

The ``mistery'' is therefore similar to the probability clock~\ref{sec:euler}: 
How is it possible that a $50/50$ probability becomes $100/0$? It is possible because precisely because the time plays no fundamental role in quantum mechanics nor in classical Hamiltonian mechanics. There is not a probability distribution for each time (or for other parameter corresponding to the symmetry group). The symmetry transformation $U'U$ is different from a stochastic process where the symmetry transformations $U'$ and then $U$ are applied, and there is no reason why it should not be different.

\section{Do the Bell inequalities hold?}
\label{sec:bell}

The Bell inequalities~\cite{bell} do
not hold---since Quantum Mechanics cannot be distinguished from a
complete statistical theory---because the assumptions of the Bell inequalities overlooked the fact that time-evolution is a stochastic process if and only if it is deterministic. As long as the time-evolution of the phase-space is a symmetry and it respects relativistic causality, there is no reasonable argument why a complete statistical theory should be a stochastic process. The whole point of the Bell inequalities is to distinguish Quantum Mechanics from a ``standard'' statistical theory, but a ``standard'' statistical theory means that the theory is completely defined by a probability distribution in a phase-space (which is the case of Quantum Mechanics and classical statistical mechanics). 

One could argue instead that the inequalities do hold, but there is an implicit assumption that the theory which is being compared to Quantum Mechanics has a time-evolution which is a stochastic process. Even in that case (see Section~\ref{sec:timetranslation}) we have that for any set of experimental results supporting relativistic Quantum Mechanics, there is a deterministic theory (and so the time-evolution is a stochastic process) which is also compatible with the same experimental results.
So, to save the Bell inequalities we would need now to find fundamental arguments against such deterministic theory. But, which arguments? Such deterministic theory is compatible with any experimental test about relativistic causality and it is not super-deterministic. These arguments would need to be somehow against the existence of pseudo-random number generators in Nature, but such generators \emph{do} exist in Nature because we humans built some of them and we are part of Nature.

To be sure, the present author does not expect that a reasonable deterministic theory will in the future replace Quantum Mechanics.
But once it is established that Quantum Mechanics is a complete statistical theory, the idea that we can rule out a reasonable deterministic theory, is also an absurd: it would imply affirming the Bayesian point of view and ruling out the Frequentist point of view. Two logical constructions can always be mutually incompatible, despite being both consistent when considered independently of each other (e.g. the Bayesian and Frequentist points of view).  In the Bayesian point of view, the probability expresses a degree of belief, and so the probability \emph{is} an entity which exists by itself. In the Frequentist point of view the root of probabilities is the absence of deterministic information that does exist somehow and is revealed through events. But if such information exists, then we cannot rule out that there is a reasonable deterministic theory which describes such information. 

In summary, either we can say that the Bell inequalities do not hold or instead, we can also say that the Bell inequalities (despite being mathematically valid inequalities) involve unrealistic assumptions which render them innocuous.

\section{Conditioned probability and constrained systems}
\label{sec:constraints}

A probability distribution can be discrete or continuous (or a linear combination of discrete and continuous probability distributions). A continuous probability distribution is a probability distribution that has a cumulative distribution function that is continuous.

In the case of continuous probability distributions, each and every single point in the phase-space has null probability. This is fortunate for the wave-function parametrization, since in the linear space of square-integrable functions ($L^2$), the point evaluation is not a continuous linear functional (that is, $L^2$ is not a reproducing kernel Hilbert space). In fact, $L^2$ is an Hilbert space of equivalence classes of functions that are equal almost everywhere (that is, up to sets with null Lebesgue measure, and null Lebesgue measure implies null probability in the context of continuous probability distributions).

But it is not obvious how to extend the wave-function parametrization to conditioned probabilities of continuous probability distributions. A conditioned probability distribution is in itself a probability distribution and so it admits a wave-function parametrization. However, the original probability distribution also admits a wave-function parametrization and the question we address now is how to relate the parametrization of the conditioned probability with the parametrization of the original probability distribution.

When deriving the continuous probability distribution from the wave-function parametrization, the value of the probability distribution at a single point of the phase-space is ambiguous and thus we cannot calculate the conditioned probability without ambiguity. Is not obvious because in the conditioned probability, we may know that an event has happened, even if the probability of such event was null (e.g. a single point in the phase space). We could argue that the conditioned probability could be only an intermediate calculation, but this would clash with the Bayesian point of view where there are only conditioned probabilities. Also from a classical mechanics point of view, a single point in the phase space does have a meaning. This ambiguity is also at the root of the need for the renormalization process in Quantum Field Theory~\footnote{Note that renormalization is about defining a multiplication of operator-valued distributions at a single point of phase-space, and the wavelet transform is related both to Hilbert spaces where the point evaluation is continuous~\cite{waveletrkhs} and to the renormalization process~\cite{battle1999wavelets}}.

The conditioned probability is a particular case of a constrained system and the ambiguity described above also appears in constrained systems in general, whenever we want to define a wave-function parametrization of a probability distribution on a subset of the phase-space defined by constraints. The constraints are from a technical point of view, a representation of an ideal by the zero number. By an ideal we mean an ideal in the algebraic sense. Regarding the normalization of the conditional probability distribution, it is automatic since the wave-function parametrization is defined independently from the ideal.

The correspondence between geometric spaces and commutative algebras is important in algebraic geometry\footnote{The correspondence between geometric spaces and commutative algebras is consequence of the Gelfand representation: there is an isomorphism between a commutative $C^*$-algebra $A$ and the algebra of continuous functions of the spectrum of $A$.}. It is usually argued that the phase space in quantum mechanics corresponds to a non-commutative algebra and thus it is a non-commutative geometric space in some sense~\cite{connesnoncommutative}.
However, after the wave-function collapse, only a commutative algebra of operators remains (see Section~\ref{sec:prob}). Thus, the phase space in quantum mechanics is a standard geometric space and the standard spectral theory (where the correspondence between geometric spaces and commutative algebras plays a main role~\cite{spectralhistory}) suffices.

%In the context of the GNS construction, we must first set the constraints to zero and only then we represent the resulting algebra  (i.e. modulo the ideal) as an algebra of operators acting on some Hilbert space. 

It suffices to constrain to zero the Casimir operators of the (eventually non-commutative) Lie algebra of constraints. This imposes the constraints without the need for the constraints to be part of the commutative algebra, only the Casimir operators are included in the commutative algebra.

Once non-determinism is taken into account, then non-commutative operators can be taken into account and the constraints are the generators of a gauge symmetry group. In case the Lie group is infinite-dimensional, there is some ambiguity in its definition~\cite{infinitelie,infinitelie}. We consider the $C^*$-algebra~\cite{realoperatoralgebras} generated by the unitary operators on an Hilbert space of the form $e^{i\int d^4 x \theta(x) G(x)}$ where $G(x)$ is a constraint and $\partial_{\mu}\theta(x)$ is a square integrable function of space-time $x$ (see also Section~\ref{sec:sft}).

Note that the algebra of observable operators already conserves the constraints (i.e. it is a trivial representation of the gauge symmetry), so the Hilbert space does not need to verify the constraints (i.e. it may be a non-trivial representation of the gauge symmetry). In fact, in many cases it would be impossible for the cyclic state of the Hilbert space to verify the constraints, as it was noted long ago:
%The separation between constraints and Hilbert space is crucial: 

\epigraph{``So we have the situation that we cannot define accurately the vacuum state. We therefore have to work with a standard ket $|S>$ which is ill-defined. One can, however, do many calculations without using the accurate conditions \emph{[vacuum verifies constraints]} and the successes of quantum electrodynamics are obtained in this way.''}{Paul Dirac (1955)~\cite{Dirac:1955uv}}

Indeed, there are some symmetries of the algebra of operators which necessarily the expectation functional  cannot have (see also~\cite{Klauder:2000gu}), since the expectation functional is a trace-class operator (the expectation of the identity operator  is 1) and its dual-space is bigger (the space of bounded operators).

For instance, consider an infinite-dimensional discrete basis $\{e_k\}$ of an Hilbert space (indexed by the integer numbers $k$) and the symmetry group generated by the transformation $e_k\to~e_{k+1}$ (translation). There is no normalized wave-function (and thus no expectation functional) which is translation-invariant, while there is a translation-invariant algebra of bounded operators (starting with the identity operator). 

%Note that We need to consider the fact that a probability theory can be defined as a particular case of a statistical theory where there is a (possibly non-commutative~\cite{parametrization}) algebra of operators and an expectation functional~\cite{whittle2000probability}.

We define gauge-fixing as comprehensive whenever it crosses all possible gauge-orbits at least once. On the other hand, we define gauge-fixing as complete whenever it crosses all possible gauge-orbits at most once, i.e. when there is no remnant gauge symmetry. The Dirac brackets require the gauge-fixing to be both comprehensive and complete, which is not possible in general due to the Gribov ambiguity~\cite{henneaux1992quantization}. 
In a non-abelian gauge-theory, the Gribov ambiguity forces us to consider a phase-space formed by fields defined on not only space but also time. This is related to the fact that in a fibre bundle (the mathematical formulation of a classical gauge theory) the time cannot be factored out from the total space because the topology of the total space is not a product of the base-space (time) and the fibre-space, despite that the total space is \emph{locally} a product space.
Thus, the Hamiltonian constraints cannot be interpreted literally, that is, as mere constraints in a too large phase-space whose ``non-physical'' degrees of freedom need to be eliminated. Moreover, this picture makes little sense in infinite-dimensions: the gauge potentials can be fully reconstructed from the algebra of gauge-invariant functions, apart from the gauge potential and its derivatives at one specific arbitrary point in space-time~\cite{wilsonloops}; thus the number of ``non-physical'' degrees of freedom would be finite at most which clearly does not match with the uncountable infinite number of constraints.

If we consider instead a commutative C*-algebra and its spectrum, such that any non-trivial gauge transformation necessarily modifies the spectrum while conserving the commutative C*-algebra  (e.g. the gauge field $A_\mu$ which is a function of space-time), then one point in the spectrum is one example of a complete non-comprehensive gauge-fixing. The gauge-fixing is non-comprehensive because the action of the gauge group on the spectrum is not transitive. Such commutative algebra has the crucial advantage that the constraints are necessarily excluded from the algebra, so that it can be used to construct a standard Hilbert space which is compatible with the constraints because the relevant operators of the commutative algebra are the ones commuting with the constraints, saving us the need to eliminate the ``non-physical'' of degrees of freedom.

In the absence of constraints, we also consider a (particular) commutative C*-algebra: the AW*-algebra.
A commutative AW*-algebra is a commutative C*-algebra whose projections form a complete Boolean algebra. Conversely, any complete Boolean algebra is isomorphic to the projections of some commutative AW*-algebra~\cite{awalgebras}. 
Therefore, the notion of probability which only arises in the absense of constraints, is as a particular case of the expectation value of an element of a commutative C*-algebra.

Thus, the Hamiltonian constraints are in fact a tool to define a probability measure for a manifold with a non-trivial topology (a principal fibre bundle for the gauge group)~\cite{gaugewhy}, because a phase-space of gauge fields defined \emph{globally} on a 4-dimensional space-time (i.e. a fibre bundle with a trivial topology, when the base space is the Minkowski space-time) produces well-defined expectation functionals for the gauge-invariant operators acting on a fibre bundle with a non-trivial topology~\cite{gaugewhy}~\footnote{Despite that the gauge potentials can only be reconstructed up to one specific arbitrary point in space-time, the probability measure for a field with support only in that point is known because it is a probability measure for a finite-dimensional space.}.
On the other hand, setting non-abelian gauge generators to zero in the wave-function would require to solve a non-linear partial differential equation with no obvious solution~\cite{gaussYM,integralYM,globalYM,dressYM}~\footnote{Note that for the case of abelian gauge theories, we could use the Hodge decomposition~\cite{Ivancevic:2008dc} of the conjugate momentum $V_\mu=\partial_\mu \phi+V_\mu'$ of the potential $A_\mu$, where $\partial^\mu V_\mu'=0$ and we set $\phi=\Delta \rho$ whenever $\phi$ appears on the left or right extreme of an operator. Then the operator $\phi$ never acts on the cyclic vector generating the Hilbert space, so that it becomes irrelevant how would $\phi$ evaluate on the cyclic vector.}.

Note that it is crucial that the C*-algebra in the gauge-fixing is commutative and it is conserved by the gauge transformations. While this is not possible in the canonical quantization, it is possible with the quantization due to time-evolution~\cite{pedro_1442442}. Note also that since only gauge-invariant operators are allowed, we must distinguish between the concrete manifold appearing in the phase-space and the family of manifolds (obtained from the concrete manifold through different choices of transition maps between local charts) to which the expectation values correspond. 

The gauge symmetry is different from anomalies.
An anomaly is a failure of a symmetry of the wave-function to be restored in the limit in which a symmetry-breaking parameter (usually introduced due to the mathematical consistency of the theory) goes to zero. We only consider symmetries of the Hamiltonian as candidate symmetries of the wave-function, since only these are respected by the time-evolution.

On the other hand, the constraints (which generate the gauge symmetry)
cannot modify the wave-functions of the Hilbert space.
Since in the case of a gauge symmetry there is no way to introduce a symmetry-breaking parameter, we can never observe an anomaly.

\section{A translation-invariant time-evolution and (classical) statistical field theory}
\label{sec:sft}

While there is a mathematically rigorous definition of classical field theory~\cite{cftmath}, so far the definition of a (classical) statistical field theory~\cite{mussardosft} is tied to the definition of a quantum field theory~\cite{Lang:1985nw}, which involves a lattice spacing necessary to regularize and renormalize the ultraviolet divergencies of the field theory. The notion of continuum limit in a discrete lattice is that for a large enough scale the predictions of the theory are independent from the type of discrete regularization used~\cite{Lang:1985nw}, thus the regularized theory is always discrete.
The regularization and renormalization are related to the decomposition of a field defined in the continuum through discrete wavelets and it is roughly the translation of the products of fields into products of wavelet components~\cite{battle1999wavelets}  (a related approach involves a semivariogram~\cite{spatialdata}). Such translation of products of fields only allows polynomial Hamiltonians; in particular when using the Fock space without regularization (which is a possible way to implement a continuous tensor product~\cite{continuoustensorp}), only quadratic Hamiltonians are allowed (i.e. for free fields).

This excludes a rigorous definition of the classical statistical version of many classical field theories (such as General Relativity), since so far there is no reason why the Hamiltonian of a classical field theory should be polynomial in the fields, not to mention the problems with Quantum Gravity~\cite{Katanaev:2005xd}. This is unacceptable: for most classical field theories, the definition of the corresponding classical statistical field theories should be straightforward, because the real-world measurements are never fully accurate.

The above is an indication that an alternative definition of Statistical Field Theory which allows the definition of non-polynomial Hamiltonians should not be too hard to find. Indeed, the essential obstruction to an infinite-dimensional Lebesgue measure is its $\sigma$-finite property (to be the countable direct sum of finite measures)~\cite{baker1991lebesgue,baker2004lebesgue}. Once we drop the $\sigma$-finite property, several relatively simple candidates exist~\cite{baker1991lebesgue,baker2004lebesgue}. In our case, we are not looking for an infinite-dimensional Lebesgue measure (no one expects the probability measure itself to be translation-invariant), but only for a translation-invariant time-evolution of the probability measure (i.e. the time-evolution is an operator, not a real number) and thus there is no reason to expect such operator to be $\sigma$-finite until it is evaluated against a probability measure when it becomes another probability measure---which cannot be translation-invariant.

The time-evolution for any quantum system is a (unitary) linear operator. This is only possible because the linear space is infinite-dimensional, this allows non-linear equations to be converted into linear equations. In the case of field theory, only free fields are allowed in Fock-spaces~\cite{Petz:1990gb} (that is, without renormalization). But a complete physical system is also a free system. If we neglect gravity, the free field associated to the free system is an orthogonal(fermion)/sympletic(boson) real representation of the Poincare group depending on whether its spin is semi-integer/integer respectively, regardless of the interactions occurring within the free system~\cite{spinstatistics,wigner}. Thus in field theory the wave-function parametrization includes a free field parametrization, as we will see in the following.

The spectrum of the boolean algebra is a subset (symmetrized(bosons)/anti-symmetrized(fermions)) 
of a power set $\mathcal{P}(S)$ of a set $S$. For an Hamiltonian involving at most second-order derivatives of the fields, the boolean algebra of a (non-free) field at an arbitrary discrete finite number $n$ of points in space is parametrized by the power set $\mathcal{P}(S)$ of a set $S=\mathbb{R}^{n*(1+2*m+(n-1)*m/2)+m}$, where $n$ is the number of space dimensions and $m$ is the number of bosonic fields (to include fermionic fields we replace $\mathbb{R}$ by the discrete set $\{0,1\}$):

\begin{align}
\phi(x_1)\cap ...\cap \phi(x_n)=(\phi,x_1)\cap ...\cap (\phi,x_n) \cap (\kappa\geq k)
\end{align}
Where $\kappa$ is the number (of free fields) operator. Thus, the boolean algebra can always be made the result of a higher-order boolean algebra where some points in space are ignored. Note that the boolean algebra of the union of two disjoint regions of space is the intersection of the two boolean algebras of the corresponding two regions of space, as it should for a field theory.

The fact that the Hamiltonian only involves local interactions allows us to introduce an *-homomorphism where a finite number of points of the continuum space is selected. Then we can do a wave-function parametrization, which allows for a non-deterministic (infinitesimal) time-evolution for these selected points. Moreover, the (deterministic or non-deterministic) time-evolution of this finite number of points can be determined independently of the probability distribution of the initial state, which would be a complex problem because it involves an infinite-dimensional Sobolev phase-space with some correlation between the points due to differentiability requirements~\cite{ringstrom2009cauchy}. This allows us to know approximately the probability distribution of the initial and final state through numerical methods for partial differential equations and regression (gaussian process regression~\cite{gpr} or statistical finite element method~\cite{statfem}, for instance).
 
Therefore, the Hamiltonian is quadratic in the creation/annihilation operators and no further regularization is needed (the free field parametrization can be considered a regularization by itself).

%The products of creation operators at the same point in space are converted into a unique creation operator whose field configuration is the product of the field configurations of the corresponding creation operators. To do this, we consider the largest commutative $C^*$-algebra of operators on the symmetric Fock-space and we impose an equivalence class on it and so the inner-product is not anymore the one from the symmetric Fock-space.

The fact that we are dealing with a \emph{commutative} algebra is key to allow the selection of only a finite number of points of the continuum space. This is only possible because we make use of the wave-function parametrization only when it is convenient, in this case only after the selection of a finite number of points of the continuum space. We can do it because the wavefunction really is just a parametrization, without a physical counterpart.

If we had assumed that there is an infinite-dimensional canonical commutation relation algebra~\cite{Petz:1990gb} from the beginning (as in most literature about Quantum Field Theory, instead of the commutative algebra we considered) then the *-homomorphism where a finite number of points of the continuum space is selected would not be possible. So our formalism includes the Fock-space (i.e. free fields), but not the other way around.

Moreover in the classical statistical field theory case where the time-evolution is deterministic\footnote{For a deterministic time-evolution the wave-function parametrization of the probability distribution for a finite number of points corresponds to the Koopman-von Neumann version of classical statistical mechanics~\cite{Sudarshan1976}.}, the wave-function parametrization allows the time-evolution operator to be unitary (and thus bounded) which is crucial to guarantee that the time-evolution of the *-homomorphism above described (and thus also the time-evolution of the expectation functional) is mathematically well defined. Without the wave-function parametrization, the selection of only a finite number of points of the continuum space is much harder already for classical field theory~\cite{finitecft}.

%Therefore, we need variables suitable to a local setting as in the classical Lagrangian field theory formalism. In particular, the partial derivative of a field $\partial_\mu \phi$  in the (global) phase-space must be replaced by an independent variable in the local phase-space, while the generators of the transformations (including the Hamiltonian and translations in space-time) must be modified accordingly.
In (quantum or classical) statistical field theory, the problem we want to solve is about a probability distribution so it is about an eigenvalue problem and diagonalizing a time-evolution operator, because the eigenfunction needs not even exist and we use ideals instead (see the previous section). On the other-hand, in classical field theory (including in numerical calculations such as the finite element method~\cite{sobolevfem,loggfem}) it is about the fields themselves and so the solution must be part of an Hilbert space (because completeness of the space is crucial for the existence proofs) and we need an alternative to the $L^2$ measure since the differential operator is unbounded with respect to the $L^2$ measure: such alternative is the Sobolev Hilbert space~\cite{sobolev}.
%The order of the Sobolev space is the minimum order such that the maximum order of the derivatives appearing in the Lagrangian are continuous (see the embedding theorem~\cite{sobolev}). Most Hamiltonians we are interested in have at most first-order derivatives in the Hamiltonian.
The cost of the free field parametrization is that we need to implement derivatives and coordinates in continuum space as an extra structure at the local level using constraints, which allows well-defined products of fields and its derivatives at the same point in the continuum space. 
For an Hamiltonian which depends on the field derivatives up to second order), the constraints are $iD_x-i\partial_x=0$ where: 
\begin{align}
%	&\phi^{(0)} i \partial_x-\phi^{(1)}=0\\
%	&\phi^{(1)} i \partial_x-\phi^{(2)}=0\\
%	&p=p_{(0)}+i \partial_x p_{(1)}+(i\partial_x)^2 p_{(2)}
    &[p_{(j)},\phi^{(k)}]=i\delta_j^k\\
    &[p_{(j)},\phi^{(2)}(x)]=i\delta_j^2\\
    &[\phi^{(j)},p_{(0)}(x)]=-i\delta_0^j\\
    &p(x)=p_{(0)}(x)\\
    &\phi(x)=\phi^{(0)}+\phi^{(1)}x+\frac{1}{2}\phi^{(2)}(x)x^2\\
    &D_x=[\partial_x,p_{(0)}(x)]\phi^{(0)}+\sum_{j=1}^{2} p_{(j-1)}\phi^{(j)}+p_{(2)}[\partial_x,\phi^{(2)}(x)]\\
    &[iD_x-i\partial_x, H]=0\\
%    &\int d\phi dy (a^\dagger(\phi,y)iD_y a(\phi,y)-i\partial_x=0\\
    %&\int d\phi dx
    %(a^\dagger(\phi,x)[iD_x,H]a(\phi,x)-[i\partial_x,a^\dagger(\phi,x)[iD_x,H]a(\phi,x)])=0\\
\end{align}

Crucially, due to the Bianchi identity and that the Hamiltonian is translation invariant, we have:
\begin{align}
    &\int d\phi dx a^\dagger(\phi,x)[[iD_x,\phi(x)],H]a(\phi,x)=-\int d\phi dx a^\dagger(\phi,x)[iDx,[H,\phi(x)]]a(\phi,x)
    &\int d\phi dx a^\dagger(\phi,x)[[iD_x,p(x)],H]a(\phi,x)=-\int d\phi dx
    a^\dagger(\phi,x)[iDx,[H,p(x)]]a(\phi,x)  
\end{align}

%For an operator $x F(\phi)$ which commutes with the constraint $iD_x-i\partial_x$, we have:
%\begin{align} 
%    &a^\dagger(\phi,x)x[iD_x-i\partial_x, F(\phi)]a(\phi,x)-a^\dagger(\phi,x)F(\phi)a(\phi,x)=0  
%\end{align}

Note that in classical Lagrangian (and Hamiltonian) Field Theory there are also derivatives of fields (through jets of modules~\cite{mathssb}), which are consistent with the commutation relations~\cite{ccrfunctions} for operator fields defined above.

The derivative of a field can be related to the derivative of the wave-function, due to the partition of unity and completness of the Hilbert space:
\begin{align}
&[\partial_x,\phi(x)]=\int d\phi dx \sum_n (a^\dagger(\phi,x))^n|0>(\partial_x \phi(x,n))<0|(a(\phi,x))^n\\
&\phi(x,n)=<0|(a(\phi,x))^n \phi (a^\dagger(\phi,x))^n|0>
\end{align}
This allows us to define the Hamiltonian dependent on only a finite number of derivatives of the field $\phi(x)$ and from it derive the time-evolution operator, relating infinite-order derivatives of the field $\phi(x)$ to infinite-order derivatives of the wave-function. Note that the smooth functions are dense in the Lebesgue square-integrable space, thus the fields (and the Hamiltonian $H$) are assumed to be smooth.

%with the (smooth, Weyl ordered) Hamiltonian density $H(\pi(x),\phi(x))$ must be related with the Euler-Lagrange equations:
%\begin{align}
%	[p,H(p,\phi^{(j)})]=&\sum_{k=0}^{2} [p_{(k)},H(p,\phi^{(j)})](i\partial_x)^k\\
%	[\phi(x),H(\pi(x),\phi(x))]=&\sum_{k=0}^{\infty} i (-1)^k \partial_\mu^k\frac{\delta H(\pi(x),\phi(x))}{\delta (\partial_\mu^k\pi(x))}
%\end{align}

%\begin{align}
	%[\pi(x),H(\pi(x),\phi(x))]=&\sum_{k=0}^{\infty} -i (-1)^k (i\partial_x)^k\frac{\delta %H(\pi(x),\phi(x))}{\delta (\partial_\mu^k\phi(x))}\\
	%[\phi(x),H(\pi(x),\phi(x))]=&\sum_{k=0}^{\infty} i (-1)^k \partial_\mu^k\frac{\delta %H(\pi(x),\phi(x))}{\delta (\partial_\mu^k\pi(x))}
%\end{align}

%Note that above, we could have defined the probability of an elementary event using a product-integral. But in the commutative case (since we are considering probability distributions, not operators), the product-integral can be replaced by the exponential of an integral~\cite{dollardproduct}, as we did. We are interested in the type II product integral~\cite{typeIandII}. The product integral is also the ``time-ordered exponential product'' used in Feynman path-integrals and the Wiener measure~\cite{johnson2000feynman}, thus it seems the natural choice for the probability distribution of random fields in general. Equivalently, we  can wavelet-transform the field in continuum space and use the (discrete) product of probability distributions for the wavelet components~\cite{mussardosft,battle1999wavelets}  (a related approach involves a semivariogram~\cite{spatialdata}).

These commutation relations imply that the field $\phi$ and its canonical conjugate $p$ commutes with a total divergence in the Hamiltonian. Thus, the divergence will not contribute to the observable quantities for probabilities which are asymptotically constant in the continuum space~\footnote{If general boundary conditions (instead of asymptotically constant probabilities) are required, we must restrict the Hamiltonian to only involve the fields and its first-order partial derivatives otherwise the formalism may be ill defined. Already in classical field theory, the Einstein-Hilbert action with the metric as a dynamical variable has second-order derivatives of the metric, which requires adding extra terms to the action for its functional derivative to be mathematically consistent in the presence of some boundary conditions~\cite{Chakraborty:2016yna}.}.

The requirement of probabilities which are asymptotically constant in the continuum space still allows for spontaneous symmetry breaking, which is about the correlation of the fields at two points in space separated by an infinite distance~\cite{yangising}. 

The resulting time-evolution operator is translation-invariant. There is still no infinite-dimensional, translation-invariant and $\sigma$-finite measure (as expected~\cite{baker1991lebesgue,baker2004lebesgue}). 
%Note that the Stone-von Neumann theorem applies to each point of the continuum space, so there is no room for ambiguities when defining the fields and its canonical conjugates at each point of the continuum space. 

We may now define Hamiltonians which are non-polynomial functions of the field $\phi(x)$ with respect to the spectral measure $d\phi(x)$, just like it happens in Quantum Mechanics (e.g. the Hamiltonian for the Hydrogen atom). Our formalism is much more powerful and general than the peculiar notion of ``continuum'' through renormalization~\cite{Lang:1985nw} also because it involves the continuum independently from renormalization and it can also be renormalized (to reproduce the standard perturbative calculations, for instance).

When defining the Hamiltonian and within each free field, we use the (symmetric) Weyl ordering~\cite{Fujii:2003ax, weylordering} due to the fact that it conserves the exponential of operators, unlike normal-ordering~\cite{normalordering}. This is an important property of the ordering, because we use often the Trotter product formula~\cite{simon2005functional,Hatano:2005gh} (e.g. in the time-evolution operator). On the other hand for the free field operators we use normal ordering, as expected in a Fock space.

\section{Conclusion}
\label{sec:conclusion}

Quantum mechanics in the ensemble interpretation, generalizes classical statistical mechanics by allowing symmetry transformations of the statistical ensemble of systems to be non-deterministic. In classical statistical mechanics any non-deterministic transformation is an external foreign element to the theory, this is unnatural for a statistical theory. Thus Quantum Mechanics in the ensemble interpretation is a natural and unavoidable generalization of classical statistical mechanics.

We showed that the wave-function is nothing else than one possible parametrization of any probability distribution. The wave-function can be described as a multi-dimensional generalization of Euler's formula, and its collapse as a generalization of taking the real part of Euler's formula.

Thus, the quantum system is entirely defined by a standard phase-space, as in classical statistical mechanics. The wave-function allows us to deal with probability theory (which involves integration) with algebraic tools. This is a common procedure in mathematics, illustrated in the following quote:

\epigraph{The fundamental notions of calculus, namely differentiation and integration, are often viewed as being the quintessential concepts in mathematical analysis, as their standard definitions involve the concept of a limit. However, it is possible to capture most of the essence of these notions by purely algebraic means (almost completely avoiding the use of limits, Riemann sums, and similar devices), which turns out to be useful when trying to generalise these concepts\emph{[...]}}
{T. Tao (2013)~\cite{integration}}

%\qitem{The challenge to give meaning to Feynman’s heuristic calculus and to define rigorously oscillatory integrals in infinite dimension, was left to mathematicians.}{}

In statistical field theory, the fact that we are dealing with a \emph{commutative} algebra is key to allow the selection of only a finite number of points of the continuum space. This is only possible because we make use of the wave-function parametrization only when it is convenient. We can do it because the wavefunction really is just a parametrization, without a physical counterpart. Even in the classical statistical field theory case where the time-evolution is deterministic, the wave-function parametrization allows the time-evolution operator to be unitary (and thus bounded) which is crucial to guarantee that the time-evolution of the expectation functional is mathematically well defined for unbounded Hamiltonians (which appear in most classical statistical field theories of interest).

A final comment about the role of time in the quantum formalism. As it was said above, in the ensemble interpretation the quantum system is entirely defined by a standard phase-space, which implies that the time plays no fundamental role in quantum mechanics. Thus, there is no ``history'', no ``trajectory'', no Lagrangian formalism and consequently there is no ``collapse''. This is consistent with the old philosophy of presentism, where time is absent from the physical reality, which was always a valid alternative to the phylosophy of eternalism, where the physical reality is a trajectory in time. In Quantum Mechanics, there are only symmetry transformations (possibly non-deterministic) of the state of the system; these transformations may be indexed by a parameter we call ``time'', but which is very different from the time of the Lagrangian formalism.

There is nothing mystical in the inconsistency of the Lagrangian formalism, since it is well known that the Lagrangian formalism cannot deal with a non-deterministic time-evolution. The crucial difference of quantum mechanics with respect to classical statistical mechanics, is a non-deterministic time-evolution, which is a natural and unavoidable generalization.

On the other hand, the mainstream literature on quantum mechanics (in its various ``interpretations'') claims that the quantum phenomena is somehow mystical and defies our everyday view of the world.
This is illustrated in the following quotes:

\epigraph{The consistent histories formalism has taught us that there are infinitely many incompatible descriptions of the world within quantum mechanics. Perhaps some simple criterion can be found to pick out one of these descriptions, by selecting one particular consistent set. Such a criterion should explain persistent quasiclassicality\emph{[...]}}{F. Dowker and A. Kent (1994)~\cite{Dowker:1994dd}}

\epigraph{It is a fundamental doctrine of quantum information science that quantum communication and quantum computation outperforms their classical counterparts. If this is to be true, some fundamental quantum characteristics must be behind better-than-classical performance of information processing tasks.\\
\emph{[...]} it will be demonstrated how quantum contextuality and violations of local realism can
be used as useful resources in quantum information applications.
}{---  \emph{M. {{\.Z}ukowski} and {\v C}. {Brukner}
    (2012)~\cite{bell_information}}}

We strongly oppose this view\footnote{Note that unexpected phenomena can always be considered mystical in the short term but not many decades later.}, there is no way in which a non-deterministic time-evolution can defy our everyday view of the world\footnote{Certainly, a non-deterministic time-evolution is compatible with the theory of classical information. The concept ``quantum information'' has nothing to do with ``science''.}.  In fact, it is the other way around: our everyday experience of the world involves a non-deterministic time-evolution; also from the point of view of mathematical complexity, many deterministic equations of motion may be extremely complex to solve while the time-evolution in quantum mechanics is a linear transformation.

In summary, in fact there is a technical problem with the definition of Field Theories (with a solution), and there is no interpretation problem with Quantum Mechanics (beyond what is reasonable to expect for any theory of Physics).

\section*{Acknowledgments}
%The author acknowledges the hospitality of the Institute of Physics at the
%University of Graz, where most of this work has been done, and of the Centro de Física Teórica
%de Partículas at the Universidade de Lisboa.
The author acknowledges Zita Marinho for convincing him to study applications of quantum methods in problems other than quantum physics, and for discussions about the use of Hilbert spaces in machine learning. 

\addcontentsline{toc}{section}{References}
\footnotesize
\singlespacing 
\bibliography{Poincare}{}
\bibliographystyle{utphysMM}
\begin{acronym}[nuMSM]
\acro{2HDM}{two-Higgs-doublet model}
\acro{ATLAS}{A Toroidal LHC ApparatuS}
\acro{BR}{Branching Ratio}
\acro{BGL}{Branco\textendash{}Grimus\textendash{}Lavoura}
\acro{BSM}{Beyond the Standard Model}
\acro{CL}{Confidence Level}
\acro{cLFV}{charged Lepton Flavor Violation}
\acro{CLIC}{Compact Linear Collider}
\acro{CMS}{Compact Muon Solenoid}
\acro{CP}{Charge-Parity}
\acro{CPT}{Charge-Parity-Time reversal}
\acro{DM}{Darkmatter}
\acro{EDM}{Electric Dipole Moment}
\acro{EFT}{Effective Field Theory}
\acro{EW}{Electroweak}
\acro{EWSB}{Electroweak symmetry breaking}
\acro{FCNC}{Flavour Changing Neutral Current}
\acro{MET}{Missing Transverse Energy}
\acro{MFV2}{Minimal Flavor Violation with two spurions}
\acro{MFV6}{Minimal Flavor Violation with six spurions}
\acro{GIM}{Glashow\textendash{}Iliopoulos\textendash{}Maiani}
\acro{GNS}{Gelfand-Naimark-Segal}
\acro{GUT}{Grand unified theory}
\acro{ILC}{International linear collider}
\acro{LEP}{Large electron\textendash{}positron collider}
\acro{LFC}{Lepton flavor conservation}
\acro{LFV}{Lepton Flavor Violation}
\acro{LHC}{Large Hadron Collider}
\acro{MFV}{Minimal flavour violation}
\acro{MIA}{Mass insertion approximation}
\acro{MSSM}{Minimal Supersymmetry Standard Model}
\acro{nuMSM}[$\nu$MSM]{minimal extension of the Standard Model by three right-handed neutrinos}
\acro{PS}{Pati-Salam}
\acro{PT}[$\mathrm{p_T}$]{transverse momentum}
\acro{QCD}{Quantum chromodynamics}
\acro{RG}{Renormalization group}
\acro{RGE}{Renormalization group equation}
\acro{SM}{Standard Model}
\acro{SUSY}{Supersymmetry, Supersymmetric}
\acro{VEV}{Vacuum expectation value}
\acro{MEG}{Muon to electron and gamma}
\acro{NP}{New Physics}
\acro{NH}{Normal hierarchy}
\acro{IH}{Inverted hierarchy}
\acro{CKM}{Cabibbo\textendash{}Kobayashi\textendash{}Maskawa}
\acro{PMNS}{Pontecorvo-Maki-Nakagawa-Sakata}

\end{acronym}
\normalsize
\onehalfspacing 
\end{document}